\documentclass[twocolumn]{aastex7}
\usepackage[whole]{bxcjkjatype}
\usepackage{amsmath}
\usepackage{hyperref}
\received{2025 July 4}
\accepted{2025 August 29}
\submitjournal{ApJ}

\begin{document}
\title{ALMA High-resolution Observation for the Transitional Disk around IRAS~04125+2902}

\author[0000-0001-6580-6038]{Ayumu Shoshi}
\affiliation{Department of Earth and Planetary Sciences, Graduate School of Science, Kyushu University, 744 Motooka, Nishi-ku, Fukuoka 819-0395, Japan}
\email[show]{shoshi.ayumu.660@s.kyushu-u.ac.jp}  

\author{Takayuki Muto}
\affiliation{Division of Liberal Arts, Kogakuin University, 1-24-2 Nishi-Shinjuku, Shinjuku-ku, Tokyo 163-8677, Japan}
\email{muto@cc.kogakuin.ac.jp}

\author[0009-0006-6416-9376]{Quincy Bosschaart}
\affiliation{Leiden Observatory, Leiden University, PO Box 9513, 2300 RA Leiden, the Netherlands}
\email{quincybosschaart@gmail.com}

\author[0000-0003-2458-9756]{Nienke van der Marel}
\affiliation{Leiden Observatory, Leiden University, PO Box 9513, 2300 RA Leiden, the Netherlands}
\email{astro@nienkevandermarel.com}

\author[0000-0002-1078-9493]{Gijs D. Mulders}
\affiliation{Instituto de Astrof\'isica, Pontificia Universidad Cat\'olica de Chile, Av. Vicu\~na Mackenna 4860, 7820436 Macul, Santiago, Chile}
\email{gijs.mulders@uc.cl}

\author[0000-0002-7951-1641]{Mitsuki Omura}
\affiliation{Department of Earth and Planetary Sciences, Graduate School of Science, Kyushu University, 744 Motooka, Nishi-ku, Fukuoka 819-0395, Japan}
\email{omura.mitsuki.362@s.kyushu-u.ac.jp}

\author[0000-0002-2062-1600]{Kazuki Tokuda}
\affiliation{Faculty of Education, Kagawa University, Saiwai-cho 1-1, Takamatsu, Kagawa 760-8522, Japan}
\affiliation{Department of Earth and Planetary Sciences, Faculty of Science, Kyushu University, 744 Motooka, Nishi-ku, Fukuoka 819-0395, Japan}
\affiliation{National Astronomical Observatory of Japan, 2-21-1 Osawa, Mitaka, Tokyo 181-8588, Japan}
\email{tokuda.kazuki@kagawa-u.ac.jp}

\author[0000-0002-0963-0872]{Masahiro N. Machida}
\affiliation{Department of Earth and Planetary Sciences, Faculty of Science, Kyushu University, 744 Motooka, Nishi-ku, Fukuoka 819-0395, Japan}
\email{machida.masahiro.018@m.kyushu-u.ac.jp}

\begin{abstract}
Recently, the youngest transiting planet was discovered around the T Tauri star, IRAS~04125+2902, in the Taurus-Auriga star-forming region. 
This system is crucial for understanding the early stages of planet formation. 
We used Atacama Large Millimeter/submillimeter Array Band 6 data to investigate the IRAS~04125+2902 system in detail. 
The dust continuum emission reveals a ring-gap transitional disk structure with an inclination of 35.6$^{\circ}$.
In addition, two-dimensional super-resolution imaging based on Sparse Modeling and the one-dimensional modeling of disk brightness distribution suggest the existence of an inner emission, which may be attributed to an inner disk, although free-free emission from the central star is not ruled out.
Furthermore, we identified the $^{12}$CO $J$=2--1 emission, and the dynamical mass of the central star is estimated to be 0.7-1.0\,$M_{\odot}$.
The asymmetry of the dust ring and the velocity distortion around the central star are, if at all, weak, suggesting that the inner disk, if it exists, is not highly inclined with respect to the outer disk.
Radiative transfer calculations of dust continuum emission suggest that the inner and the outer disk may be misaligned by $\sim$10$^\circ$, which may be confirmed in future observations with higher resolution and sensitivity.
Our results suggest that IRAS~04125+2902 is a dynamically complex system, where the binary orbit, outer disk, inner disk, and planetary orbit are mutually misaligned, providing insight into the early orbital evolution of young systems.
\end{abstract}

\keywords{Protoplanetary disks(1300), Exoplanet migration(2205), Radio interferometry(1346)}

\section{Introduction}\label{sec:introduction}
Planet formation is considered to occur in the protoplanetary disk after the mass accretion phase ends \citep[e.g.,][]{Hayashi_1985, Shu_1987}. 
Recent observations by several telescopes, including the Atacama Large Millimeter/submillimeter Array (ALMA) and VLT/SPHERE, have revealed protoplanets with possibly circumplanetary disks within protoplanetary disks around young stars, such as PDS70, ABAur, and HD169142 \citep[][]{Keppler_2019,Benisty_2021,Currie_2022,Hammond_2023}. 
Furthermore, observations of CO isotopic lines toward protoplanetary disks reveal signatures of planet-disk interaction, including gas cavities and kinks \citep[][]{vanderMarel_2016,Peres_2015, Pinte_2018}.
Thus, detailed characterizations of the physical structure and dynamics of protoplanetary disks with planets, including those containing exoplanets, are crucial for understanding the formation and evolution of planetary systems.

In this study, we focus on IRAS~04125+2902 in the Taurus-Auriga star-forming region. 
IRAS~04125+2902 is a T Tauri star with the spectral type of M1, located at 160.1\,pc from the Sun \citep[][]{Gaia_2023}. 
The mass, effective temperature and the luminosity are estimated to be 0.5-0.7\,$M_{\odot}$, $\sim$3765\,K and 0.4-0.5\,$L_{\odot}$, respectively, by the analyses of the spectral energy distribution (SED) of the system \citep[e.g.][]{Espaillat_2015, Testi_2022, Barber_2024}.
This system was observed by NASA's Transiting Exoplanet Survey Satellite (TESS) and confirmed to host a Jovian-mass planet with an orbital period of 8.83\,days \citep[][]{Barber_2024}.
Multiple indicators suggest that the star is extremely young, with an estimated age of 3.0$\pm$0.4\,Myr \citep[][]{Barber_2024,Luhman_2025}, and the transiting planet, IRAS~04125+2902~b, is currently the youngest known transiting exoplanet.
The radius of the planet ($\sim$0.96\,$R_{\rm Jupiter}$) suggests that it was probably not formed in-situ, but rather formed at a larger orbital distance and subsequently migrated inward.

One of the most remarkable features of this system is that the orbit of IRAS~04125+2902~b is misaligned with the protoplanetary disk.
\citet{Espaillat_2015} revealed a face-on transitional disk by the Submillimeter Array (SMA) observation with a spatial resolution of $\sim$0$\farcs$29 ($\sim$47\,au).
In contrast, the detection of the planet transit indicates that its orbit is close to edge-on.
This configuration presents a unique opportunity to study the dynamical processes that lead to the formation of inclined planetary orbits during the early stages of planet formation.
The same SMA observations of the $^{12}$CO $J$=3--2 line in \citet{Espaillat_2015} failed to detect any emission associated with this system due to the insufficient sensitivity.

Recently, Bosschaart et al. (submitted) presented a survey of a large sample of protoplanetary disks, including one around IRAS~04125+2902, using the ALMA Band 6 data with a resolution of $\sim 0\farcs12$ (19\,au). 
Still, the dataset for IRAS~04125+2902 has not been analyzed in detail.
In this paper, we use the same observation data, including dust continuum and $^{12}$CO $J$=2--1, to investigate the dust and gas structures in greater detail.
This paper is organized as follows.
In Section~\ref{sec:obs_imaging}, we describe the data reduction and our imaging methods, CLEAN and Sparse Modeling (SpM).
Section~\ref{sec:results} presents the results of the continuum and $^{12}$CO emissions, where we estimate the physical quantities of the IRAS~04125+2902 system.
In Section~\ref{sec:disk_modeling}, we explore an axisymmetric disk brightness distribution model for further investigations of the disk structures.
In Section~\ref{sec:discussion}, we discuss possible inner disk signatures.
Finally, Section~\ref{sec:summary} is for the conclusions.

\begin{figure*}[ht]
    \centering
    \includegraphics[width=\linewidth]{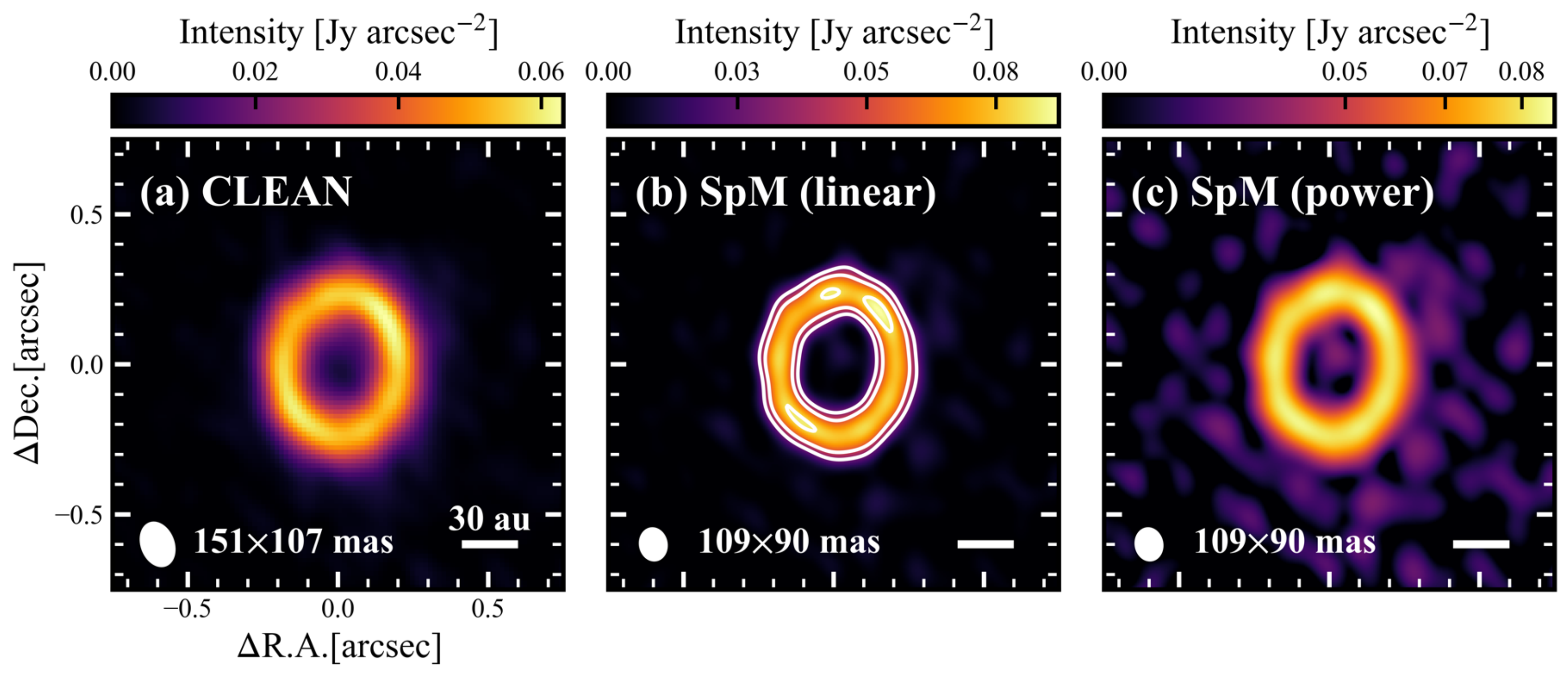}
    \caption{
    ALMA 1.3 mm (Band 6) dust continuum around the IRAS~04125+2902 system.
    (a) CLEAN image with Briggs weighting (\texttt{robust}=0.5). 
    The white ellipse indicates the synthesized beam.
    (b) SpM image with a color scale following a linear law with a scaling exponent of 1.0. 
    The white ellipse represents the effective spatial resolution estimated using the point-injection method.
    The white contours show the continuum emission at 0.03, 0.05, and 0.08\,Jy\,arcsec$^{-2}$.
    (c) Same as panel (b), but with the color scale given by a power law with a scaling exponent of 0.5. 
    }
    \label{fig:continuum}
\end{figure*}

\begin{figure*}[ht]
    \centering
    \includegraphics[width=0.88\linewidth]{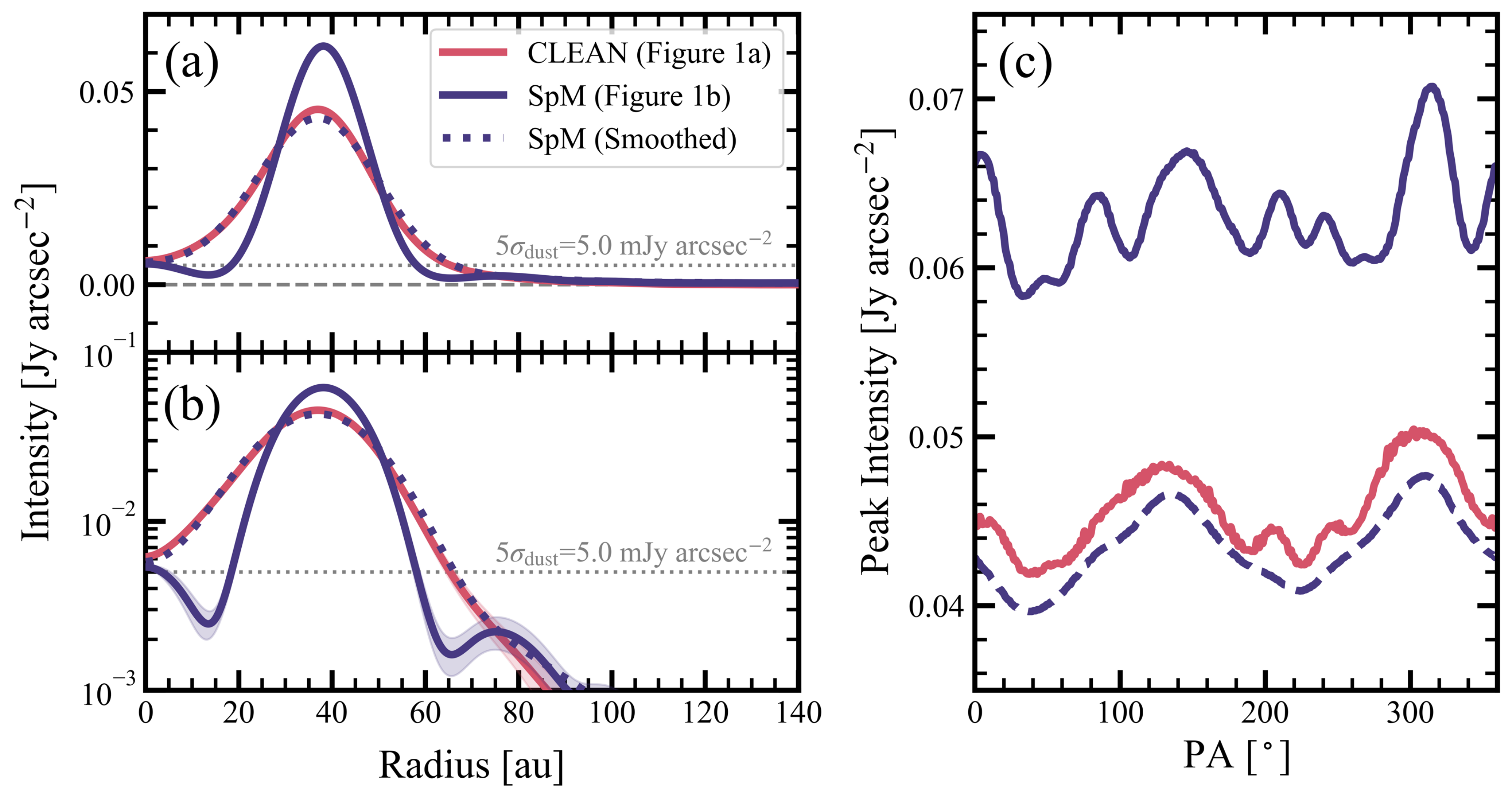}
    \caption{
    (Left) Radial intensity profile averaged over the full azimuthal angle, shown on a linear scale in the top panel and a logarithmic scale in the bottom one.
    The profile is linearly interpolated onto radial grid points spaced by 0.1\,au using \texttt{interpolate.interpld} from the \texttt{SciPy} module.
    The light colored ribbon represents the error of the mean at each radius.
    (Right) Peak intensity profile in the azimuthal direction after their brightness distributions were deprojected using $i_{\rm disk}$ and PA.
    The dotted violet lines show the profiles of the SpM image smoothed with the CLEAN beam by the CASA task \texttt{imsmooth}.
    }
    \label{fig:cont_profile}
\end{figure*}

\section{ALMA Observation \& Imaging}\label{sec:obs_imaging}
\subsection{Data Reduction}\label{subsec:reduction}
We used a subset of ALMA archival Band 6 data from the Cycle 9 project \#2022.1.01302.S (PI: Gijs D. Mulders).
The dataset includes observations of 26 protoplanetary disks (including three binary systems) located in Taurus, Lupus, Chamaeleon, and Ophiuchus.
Observations were conducted on May 21st and May 23rd, 2023, using the C-6 and C-7 configurations, with baselines ranging from 27\,m to 3638\,m and an on-source integration time of approximately 30 minutes.

The dataset consists of two spectral windows for dust continuum and two for molecular line emissions (for details, see \S\ref{subsec:imaging_cont} and \S\ref{subsec:imaging_line}).
The raw data were calibrated using the ALMA pipeline within the Common Astronomy Software Applications package \citep[CASA;][]{CASA_2022}.
An overview of the project results will be presented in Bosschaart et al. (submitted).
In this paper, we focus on the IRAS~04125+2902 dataset to conduct a detailed investigation of the protoplanetary disk structure and gas distribution.

\subsection{Imaging Dust Continuum}\label{subsec:imaging_cont}
\subsubsection{CLEAN}\label{subsubsec:imaging_clean}
We used two SPWs with the central frequencies of 218 and 233\,GHz and the bandwidth of 2\,GHz each for imaging dust continuum.
The continuum data were imaged using the CASA task \texttt{tclean} of CASA version 6.1.0.
We consistently employed multi-frequency synthesis \citep[\texttt{nterm}=2;][]{Rau_2011} and the Cotton-Schwab algorithm \citep[][]{Schwab_1984}, using Briggs weighting with \texttt{robust}=0.5.

We also adopted two rounds of self-calibration to correct gain errors and improve the signal-to-noise ratio (SNR).
In the first stage, we applied phase self-calibration (\texttt{calmode=p}) with an integration time equal to the on-source time (OST).
Next, we performed amplitude and phase self-calibration (\texttt{calmode=ap}) with an integration time of OST/5.
The final continuum image shows an improvement in SNR by a factor of 1.5 after self-calibration (the final peak SNR=62.3).
The beam size is $0\farcs151\times0\farcs107$ with a position angle (PA) of 21.0$^\circ$, and the RMS noise level ($\sigma_{\rm dust}$) is 1.0\,mJy\,arcsec$^{-2}$ ($\sim$19\,$\mu$Jy\,beam$^{-1}$).

\subsubsection{Sparse Modeling}\label{subsubsec:imaging_spm}
We utilized SpM imaging with the self-calibrated dataset to reconstruct a super-resolution image surpassing the CLEAN image in quality.
We used the SpM imaging software \texttt{PRIISM}\footnote{\url{https://github.com/tnakazato/priism}} \citep[][]{Nakazato_2020,Nakazato_2020b} version 0.11.5 on CASA version 6.1.0 to perform $\ell_1$+TSV imaging with a cross-validation (CV) scheme, following the methodology of \citet{Yamaguchi_2024} and \citet{Shoshi_2025}.
This imaging technique minimizes a cost function, which consists of a chi-squared error term representing the difference between the observed visibility and the visibility model, which is derived from the model image via Fourier transformation, along with two additional regularization terms: the $\ell_1$-norm and the total squared variation TSV.

The first regularization term, the $\ell_1$-norm, enhances image sparsity by preserving the total flux while suppressing low-intensity noise \citep[][]{Honma_2014}.
The second regularization term, TSV, ensures a smooth brightness distribution by minimizing the squared differences between adjacent pixels \citep[][]{Akiyama_2017, Kuramochi_2018}.
The hyperparameters $\Lambda_l$ and $\Lambda_{tsv}$, associated with the $\ell_1$-norm and TSV, control the relative weighting of the regularization terms with respect to the observations.
We explored a broad range of values for the two hyperparameters $(\Lambda_l, \Lambda_{tsv})$. 
We selected the optimal combination for each image by minimizing the cost function using the 10-fold CV approach \citep[][]{Yamaguchi_2021}.
As a result, we obtained the final model image with $(\Lambda_l, \Lambda_{tsv})=(10^5, 10^{13})$.

To evaluate the effective spatial resolution of the model image, we adopted the `point-source injection' method, as described in \citet{Yamaguchi_2021}. 
We injected an artificial point source, whose total flux was set to 5\% of the target's total flux, into an emission-free region north of the central star at a distance within the maximum recoverable scale ($\sim$2.1\,arcsec).
The effective spatial resolution is $0\farcs109\times0\farcs090$ with a PA of 10.4$^\circ$, which is approximately 1.5 times finer than that of the CLEAN image.

\subsection{Imaging Molecular Line Emission}\label{subsec:imaging_line}
The dataset for line emissions consists of two SPWs with bandwidths of 0.23\,GHz and central frequencies of 219.99 and 230.53\,GHz.
For line imaging, we used \texttt{tclean} with Briggs weighting (\texttt{robust}=0.5) and the \texttt{multi-scale} algorithm.
Mask regions for CLEAN imaging were automatically determined using the \texttt{auto-multithresh} scheme \citep[][]{Kepley_2020}.

The SPW centered at 219.99\,GHz does not cover $^{13}$CO and C$^{18}$O $J$=2--1 lines but covers SO $6_5$--$5_4$. 
However, the analysis of SO $6_5$--$5_4$ using the SPW centered at 219.99\,GHz did not reveal any significant emission around IRAS~04125+2902. 
For the SPW centered at 230.53\,GHz, we detected $^{12}$CO $J$=2--1 line emission.
The synthesized beam size in the $^{12}$CO $J$=2--1 line image was $0\farcs155\times0\farcs106$ with a PA of 27.4$^\circ$.
The RMS noise level ($\sigma_{\rm CO}$) at a velocity resolution of 0.2\,km\,s$^{-1}$ was 3.8\,mJy\,beam$^{-1}$.

\begin{figure*}[ht]
    \centering
    \includegraphics[width=\linewidth]{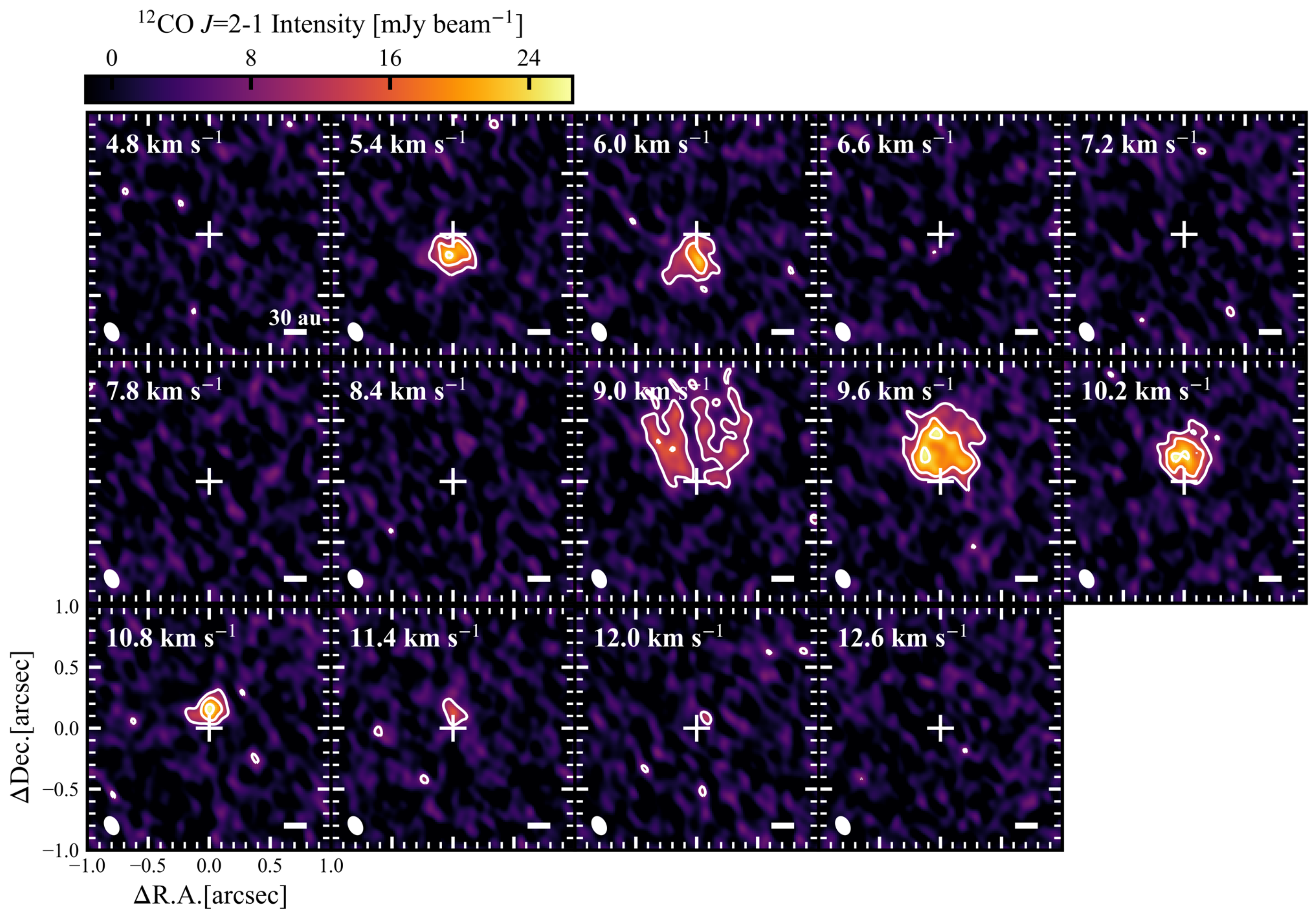}
    \caption{
    Velocity-channel maps of $^{12}$CO $J$=2--1 emission toward IRAS~04125+2902 with a velocity resolution of 0.6\,km\,s$^{-1}$.
    The central velocity $v_{\rm cent}$ is to be 7.7\,km\,s$^{-1}$.
    The white cross represents the position of IRAS~04125+2902.
    The white contours show the $^{12}$CO $J$=2--1 emission for 8.0, 16.0, and 24.0\,mJy\,beam$^{-1}$.
    The white ellipse in the lower left corner of each panel represents the synthesized beam of $0\farcs155\times0\farcs106$ (27\,au$\times$17\,au) with a PA of 27.4$^\circ$.
    }
    \label{fig:channel}
\end{figure*}

\section{Results}\label{sec:results}
\subsection{1.3 mm Dust Continuum}\label{subsec:dust}
Figure~\ref{fig:continuum} shows the 1.3\,mm dust continuum images generated by CLEAN and SpM.
We identified the transitional disk with a clear ring structure in both images.

We first fitted the ring shown in the SpM image with an ellipse and derived the inclination angle $i_{\rm disk}$ and PA, following the method described in \citet{Yamaguchi_2021} and \citet{Shoshi_2024}.
The values of $i_{\rm disk}$ and PA$_{\rm disk}$ were estimated to be 35.6$\pm$0.2$^\circ$ and 175.1$\pm$0.5$^\circ$, respectively.
This suggests that the transitional disk around IRAS~04125+2902 is nearly face-on and misaligned with respect to both the binary orbit of 2MASS~J04154269+2909558 and the orbital plane of the transiting planet IRAS~04125+2902~b, both of which have inclination angles of $\sim$90$^\circ$ \citep[][]{Barber_2024}.

Left panels of Figure~\ref{fig:cont_profile} show the azimuthally averaged radial intensity profiles of the CLEAN and SpM images deprojected using $i_{\rm disk}$ and PA$_{\rm disk}$.
We applied the Gaussian fitting using the \texttt{optimize.leastsq} function in the \texttt{Scipy} module, and the width (FWHM) of the ring in the SpM image was estimated to be 21.7$\pm$7.4\,au.
We conservatively estimated that the uncertainty was the standard deviation derived from the major-axis size of the effective spatial resolution $\theta_{\rm eff}$.

The radial profile of intensity indicates that the gap is not empty.  
The profile of the CLEAN image shows dust emission above 5$\sigma_{\rm dust}$ level ($\sigma_{\rm dust}$=1.0\,mJy\,arcsec$^{-2}$).
The SpM image of Figure~\ref{fig:continuum}(c) and its profile indicate the compact weak emission within the gap.

To investigate the azimuthal variations of the ring, we deprojected the images and traced the peak of the radial intensity profiles at different PAs.
As shown in Figure~\ref{fig:cont_profile}(c), we saw weak azimuthal variations in both SpM and CLEAN images. 
We will discuss the azimuthal asymmetry in more detail in \S\ref{subsec:comparison} and \ref{subsec:shadow}.

We then applied the curve-growth method \citep[e.g.,][]{Ansdell_2016} to measure the flux density $F_\nu$ and the disk radius $R_{95\%, {\rm dust}}$, corresponding to the radius enclosing 95\% of $F_\nu$.
In the SpM image, we estimated $F_\nu$=13.0$\pm$1.3\,mJy and $R_{95\%, {\rm dust}}$=52.8$\pm$7.4\,au.
The uncertainties in $F_\nu$ and $R_{95\%, {\rm dust}}$ were attributed to a 10\% absolute calibration error in ALMA observations and the standard deviation derived from the major axis of the effective spatial resolution.
The value of $R_{95\%, {\rm dust}}$ was also consistent with the estimation in \citet{Espaillat_2015} and Bosschaart et al. (submitted).
Based on the flux density of the SMA observations in \citet{Espaillat_2015} ($\sim$41.0$\pm$1.0\,mJy at 345\,GHz), we estimated the spectral index of $\sim$2.7.

Assuming that the emission is optically thin and comes only from a thermal component, we estimated the dust mass using,
\begin{align}
    M_{\rm dust}=\frac{F_\nu d^2}{\kappa_\nu B_\nu\left(T_{\rm dust}\right)},\label{eq:dust_mass}
\end{align}
where $F_\nu$ is the flux density, $d$ is the distance $d$=160.1\,pc, $\kappa_\nu$ is the absorption coefficient adopted as $\kappa_\nu$=2.3\,cm$^2$\,g$^{-1}$ \citep[][]{Beckwith_1990}, $B_\nu\left(T_{\rm dust}\right)$ is the Planck function at the observed frequency of $\sim$225\,GHz, and $T_{\rm dust}$ is the dust temperature.
We adopted $T_{\rm dust}$=20\,K, which is the standard value for Class II systems \citep[][]{Pascucci_2016}.
We then obtained $M_{\rm dust}\sim$3.0$\times$10$^{-5}$\,$M_\odot$ ($\sim$10.1\,$M_\oplus$).
Note that this estimate of dust mass relies on the assumption of optically thin emission and depends heavily on the assumed opacity and temperature.
In \S\ref{subsec:shadow}, we check the validity of the optically thin assumption using a simple radiative transfer calculation.  
For opacity, there is uncertainty of one order of magnitude.
Yet, we consider that the derived dust mass ($\sim10.1\,M_{\oplus}$) is smaller than or comparable to the mass of the transiting planet ($\sim 90~M_{\oplus}$).
The physical quantities of the dust disk are summarized in Table~\ref{table:quantities}.

\begin{figure}
    \centering
    \includegraphics[width=\linewidth]{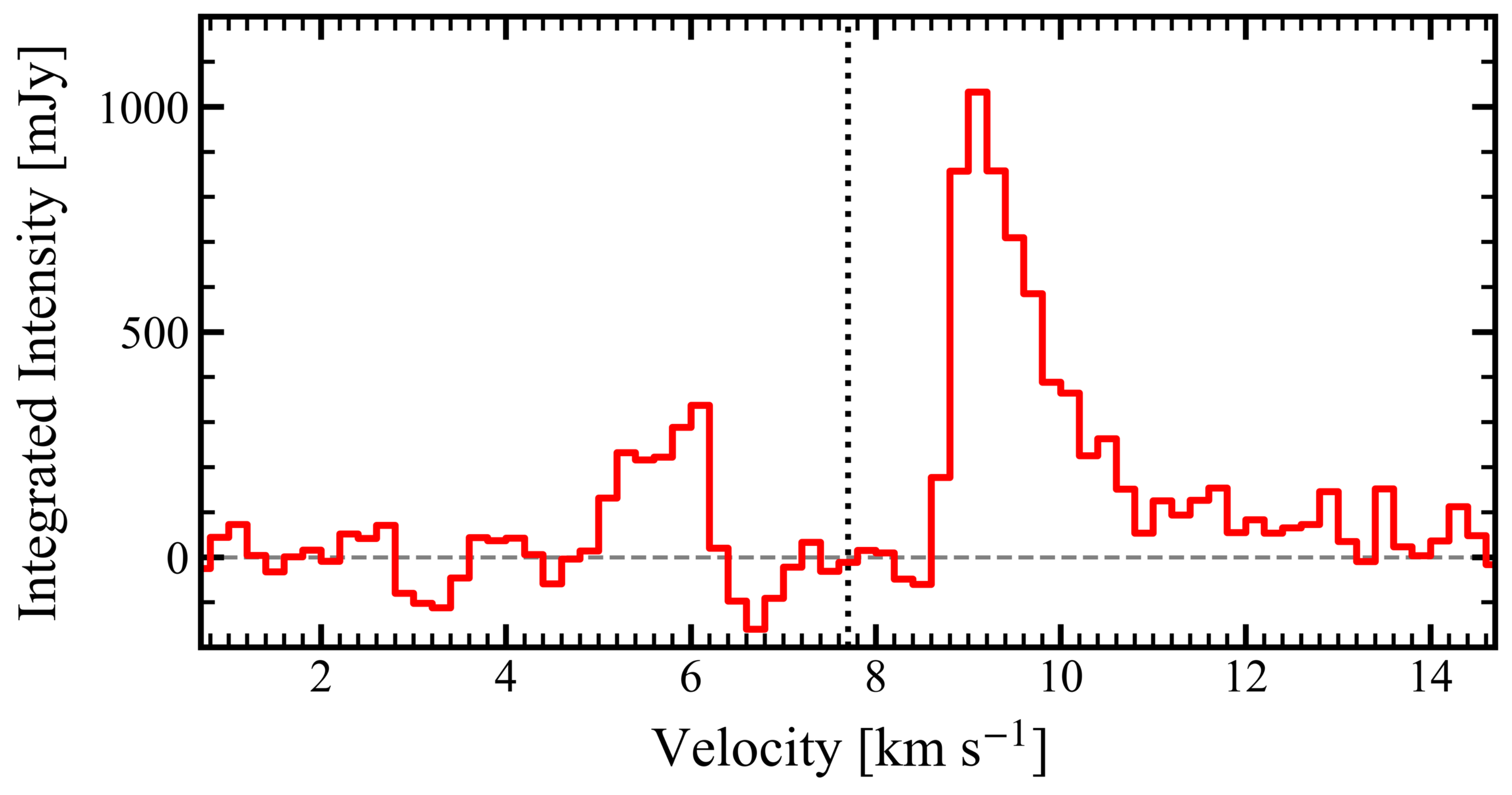}
    \caption{
    Spectrum of $^{12}$CO $J$=2--1 emission integrated within the region in a radius of 1\,arcsec from the central star observed by ALMA.
    The black dotted line shows the central velocity of $v_{\rm cent}$=7.7\,km\,s$^{-1}$.
    }
    \label{fig:spectra}
\end{figure}

\begin{figure*}[ht]
    \centering
    \includegraphics[width=\linewidth]{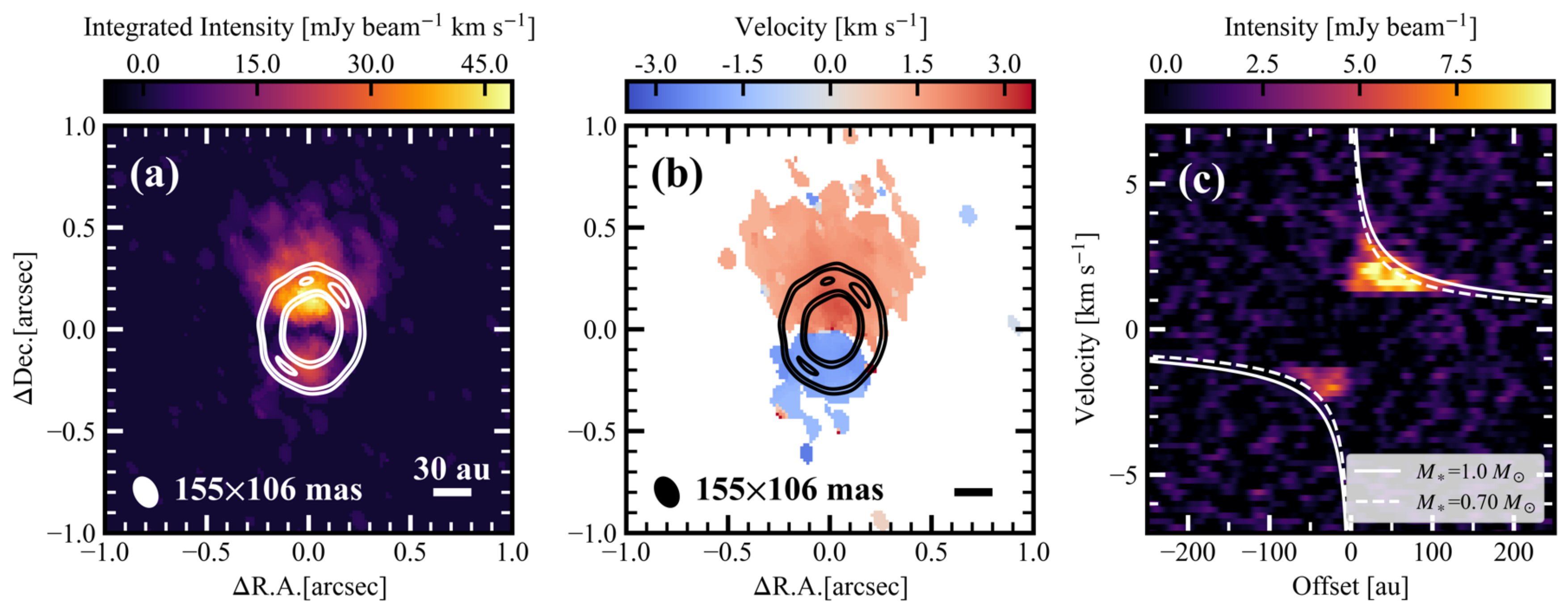}
    \caption{
    (a) Velocity-integrated intensity (moment 0) map of $^{12}$CO $J$=2--1 emission with intensities  greater than 3\,$\sigma_{\rm CO}$ ($\sigma_{\rm CO}$=3.8\,mJy\,beam$^{-1}$) in the velocity ranges of 4.6--7.6\,km\,s$^{-1}$ and 7.8--12.8\,km\,s$^{-1}$.
    (b) Velocity-field (moment 1) map of $^{12}$CO $J$=2--1 emission.
    The plot area is limited to the region larger than 4.5$\times$10$^{-3}$\,arcsec$^2$.
    The velocity of 0\,km\,s$^{-1}$ corresponds to the central velocity $v_{\rm cent}$=7.7\,km\,s$^{-1}$.
    The contours in panels (a) and (b) represent the 1.3\,mm continuum emission from the SpM image shown in Figure~\ref{fig:continuum}(b), with contour levels of 0.03, 0.05, and 0.08\,Jy\,arcsec$^{-2}$.
    (c) Position-velocity diagram along the position angle (PA) of the dust disk, from south to north.
    The offset of 0\,au corresponds to the position of the star.
    The white dashed and solid curves represent the Keplerian rotation for the stellar mass $M_\ast$=0.7\,$M_\odot$ and 1.0\,$M_\odot$, respectively.
    Note that these curves are corrected for the inclination angle of dust disk $i_{\rm disk}$.
    }
    \label{fig:coline}
\end{figure*}

\begin{figure*}[ht]
    \centering
    \includegraphics[width=\linewidth]{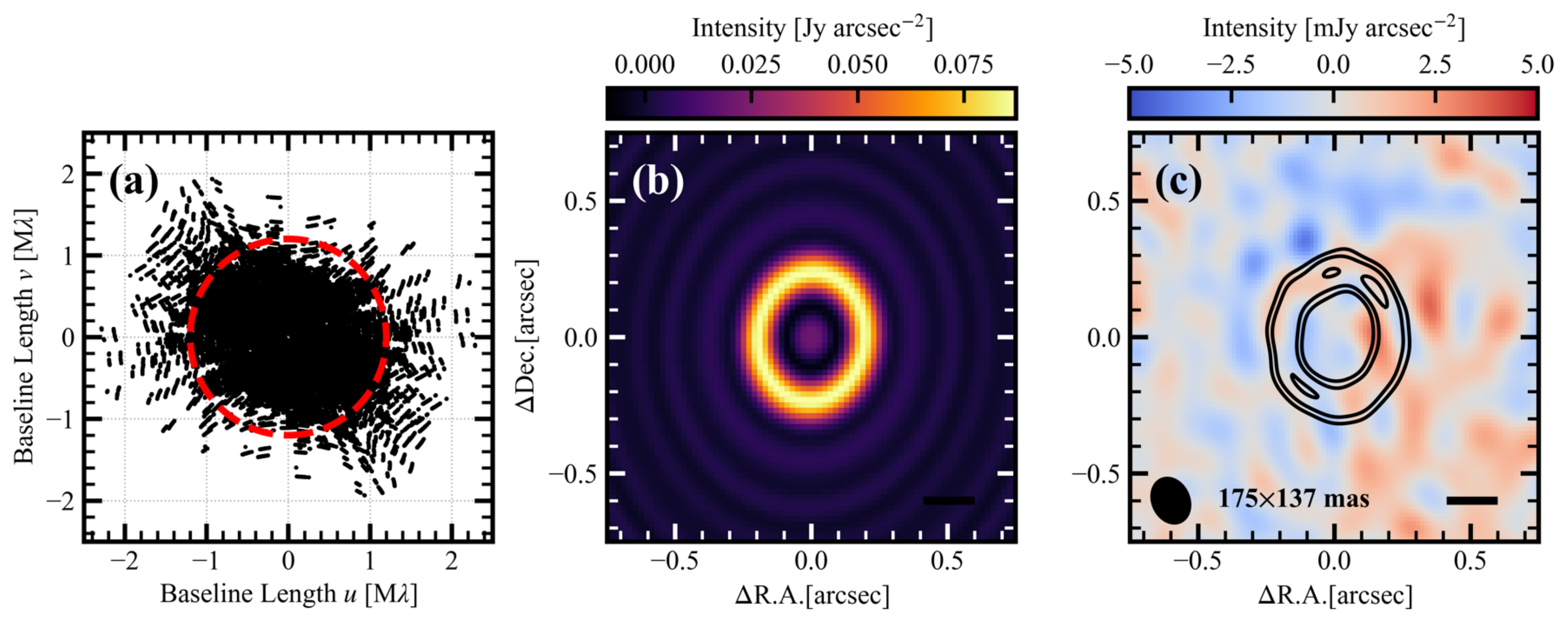}
    \caption{
    (a) $uv$ coverage of the ALMA observation toward IRAS~04125+2902 with the configurations of C-6 and C-7.
    The region within the red dashed circle, with a radius of 1.2\,M$\lambda$, indicates the dataset used for the axisymmetric disk modeling described in \S\ref{subsec:vis_selection}.
    (b) Disk model image created using the best-fit values of $\cos i$ and PA, assuming an axisymmetric disk.
    (c) Residual map generated by applying \texttt{tclean} to the dataset obtained after subtracting the model visibility from the observed visibility.
    The black contours represent the 1.3\,mm continuum emission from the SpM image shown in Figure~\ref{fig:continuum}(b), with contour levels of 0.03, 0.05, and 0.08\,Jy\,arcsec$^{-2}$.
    }
    \label{fig:model}
\end{figure*}

\begin{figure*}[ht]
    \centering
    \includegraphics[width=0.95\linewidth]{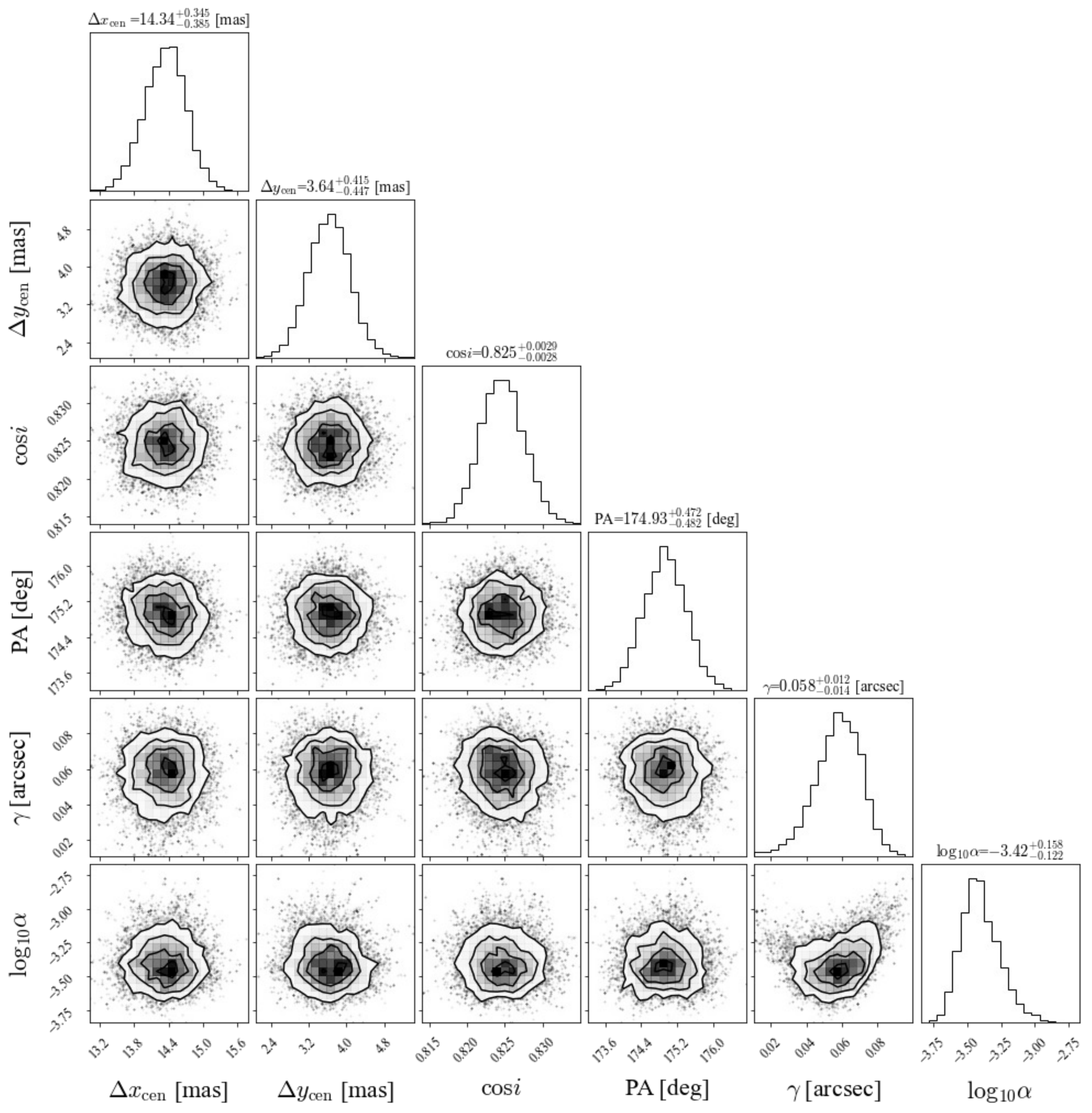}
    \caption{
    Corner plot of the posterior distribution for six parameters ($\Delta x_{\rm cen},\,\Delta y_{\rm cen},\,\cos i,\,{\rm PA},\,\gamma,\,\alpha$) obtained from the axisymmetric disk modeling applied to the visibility data of the dust continuum emission around IRAS~04125+2902.
    The most probable parameter values are displayed at the top of the histogram panels.
    }
    \label{fig:corner}
\end{figure*}

\begin{table*}[t]
    \centering
    \caption{Properties of the dust and gas disk}
    \label{table:quantities}
    \begin{tabular}{ccccccccc}
    \hline
    \multicolumn{5}{c}{Dust disk} & \multicolumn{4}{c}{Gas disk of $^{12}$CO $J$=2--1} \\
    ${\rm PA_{disk}}$ & $i_{\rm disk}$ & $F_\nu$ & $R_{95\%, \rm dust}$ & $M_{\rm dust}$ & $v_{\rm cent}$ & $F_{^{12}\rm CO}^{\rm corr}$ & $R_{95\%, \rm gas}$ & $M_{\rm H_2}^{\rm corr}$ \\
    $\left[^\circ\right]$ & $\left[^\circ\right]$ & $\left[{\rm mJy}\right]$ & $\left[{\rm au}\right]$ & $\left[M_\odot\,\left(M_\oplus\right)\right]$ & $\left[{\rm km}\,{\rm s}^{-1}\right]$ & $\left[{\rm mJy\,km\,s^{-1}}\right]$ & $\left[{\rm au}\right]$ & $\left[M_\odot\right]$ \\
    \hline
    $175.1\pm0.5$ & $35.2\pm0.2$ & $13.0\pm1.3$ & $52.8\pm7.4$ & $3.0\times10^{-5}$ (10.1) & $7.7$ & $702.2\pm5.9$ & $98.4\pm10.5$ & $\gtrsim$0.2-5.8$\times10^{-5}$ \\
    \hline
    \end{tabular}
\end{table*}

\subsection{\texorpdfstring{$^{12}$CO $J$=2--1 Emission}{12CO J=2--1 Emission}}\label{subsec:12co}
We detected $^{12}$CO $J$=2--1 emission around IRAS~04125+2902.
Figure~\ref{fig:channel} shows the velocity-channel map of $^{12}$CO $J$=2--1 emission in the velocity range of 4.8 to 12.6\,km\,s$^{-1}$, with a velocity resolution of 0.6\,km\,s$^{-1}$. 
The velocity resolution was adjusted using the CASA task \texttt{imrebin}.
The emissions extended to the south and north, corresponding to the blueshifted and redshifted components, respectively.
Some channels in Figure~\ref{fig:channel} show that the $^{12}$CO emission could be affected by foreground absorption.
We detected several components with intensities exceeding 16\,mJy\,beam$^{-1}$ in the redshifted emission, suggesting that the gas distribution could be locally enhanced.
We measured the $^{12}$CO $J$=2--1 spectrum integrated within the region in a radius of 1\,arcsec from the central star, shown in Figure~\ref{fig:spectra}.
The spectrum exhibits a blueshifted peak at 6.2\,km\,s$^{-1}$ and a redshifted peak at 9.2\,km\,s$^{-1}$.
We estimated that the systemic velocity was $\sim$7.7\,km\,s$^{-1}$. 

Figures~\ref{fig:coline} (a) and (b) present the velocity-integrated intensity (moment 0) map and the velocity-field (moment 1) map of $^{12}$CO $J$=2--1 emission, respectively.
The blueshifted and redshifted emissions are distributed over a broad region spanning 80--100\,au from the central star in the velocity range of $v_{\rm cent}\pm$5\,km\,s$^{-1}$.
Compared to the dust emission shown in Figure~\ref{fig:continuum}, the $^{12}$CO $J$=2--1 emission traces the Keplerian rotation of the gas disk around IRAS~04125+2902, with its rotation axis nearly aligned with the normal to the dust disk.
Figure~\ref{fig:coline} (c) shows the position-velocity diagram along the major axis of the dust disk from south to north.
The Keplerian curve for a stellar mass $M_\ast$=0.7\,$M_\odot$ \citep[][]{Barber_2024} agrees with the edge of the blueshifted emission, while that for $M_\ast$=1.0\,$M_\odot$ matches the edge of the redshifted emission. 
The blueshifted emission is more spatially confined than the redshifted emission, suggesting that the foreground absorption of $^{12}$CO $J$=2--1 emission should be considered for the blueshifted component.
Therefore, the stellar mass $M_\ast$ could be closer to 1.0\,$M_\odot$ than 0.7\,$M_\odot$ or greater than 1.0\,$M_\odot$.

Following the method of \citet{Dunham_2014}, we derived a lower limit to the gas mass from $^{12}$CO $J$=2--1 emission shown in Figure~\ref{fig:coline}.
We first calculated the H$_2$ column density $N_{{\rm H}_2}$, assuming that the gas components are optically thin and follow local thermal equilibrium (LTE) conditions.
The H$_2$ column density $N_{{\rm H}_2}$ can be estimated using
\begin{align}
    N_{{\rm H}_2}&=f\left(J,T_{\rm ex}, X_{\rm CO}\right)I_{\rm CO},\label{eq:column_density}\\
    f\left(J,T_{\rm ex}, X_{\rm CO}\right)&=X_{\rm CO}^{-1}\frac{3k}{8\pi^3\nu\mu^2}\frac{2J+1}{J+1}\frac{Q\left(T_{\rm ex}\right)}{g_J}e^{\frac{E_{J+1}}{kT_{\rm ex}}},\label{eq:factor}
\end{align}
where $I_{\rm CO}$ is the integrated intensity in K\,km\,s$^{-1}$ measured in the region exceeding  3\,$\sigma_{\rm CO}$, $k$ is the Boltzmann constant, $\nu$ is the rest frequency of $^{12}$CO $J$=2--1, $\mu$ is the magnetic dipole moment, $J$ is the lower rotation state, $X_{\rm CO}$ is the $^{12}$CO abundance relative to H$_2$, and $T_{\rm ex}$ is an excitation temperature.
In these equations, $f\left(J,T_{\rm ex}, X_{\rm CO}\right)$ is a function of the quantum number, and $Q\left(T\right)$ is the partition function.
Assuming $X_{\rm CO}$=10$^{-4}$ \citep[e.g.,][]{Frerking_1982} and $T_{\rm ex}$=50\,K \citep[][]{Ginsburg_2011}, we estimated $f\left(J,T_{\rm ex}, X_{\rm CO}\right)$=7.2$\times$10$^{18}$\,cm$^{-2}$\,(K\,km\,s$^{-1}$)$^{-1}$ and derived the averaged H$_2$ column density $N_{{\rm H}_2}$=2.2$\times$10$^{20}$\,cm$^{-2}$.
Note that the values of $f\left(J,T_{\rm ex}, X_{\rm CO}\right)$ and $N_{{\rm H}_2}$ at $T_{\rm ex}$=20\,K were smaller by a factor of 1.5 than those at $T_{\rm ex}$=50\,K.
Then, we calculated the gas mass $M_{{\rm H}_2}$ as,
\begin{align}
    M_{{\rm H}_2}=\mu_{{\rm H}_2} m_{\rm H} N_{{\rm H}_2} \Delta S,\label{eq:gas_mass}
\end{align}
where $\mu_{{\rm H}_2}$ is the mean molecular weight per hydrogen molecule \citep[$\mu_{{\rm H}_2}$=2.8,][]{Evans_2022}, $m_{\rm H}$ is the mass of a hydrogen atom, and $\Delta S$ is the area of a pixel with more than 3\,$\sigma_{\rm CO}$.
The gas mass $M_{{\rm H}_2}$ estimated from $^{12}$CO $J$=2--1 emission is 1.3$\times$10$^{-6}$\,$M_\odot$.
In the case of adopting the simple CO-to-H$_2$ conversion factor of 2.0$\times$10$^{20}$cm$^{-2}$\,(K\,km\,s$^{-1}$)$^{-1}$ \citep[][]{Balatto_2013}, we derived $M_{{\rm H}_2}$=3.5$\times$10$^{-5}$\,$M_\odot$, indicating that $M_{{\rm H}_2}$ would be lower than or comparable to the dust mass $M_{\rm dust}$ calculated in \S\ref{subsec:dust}.
We note that the value of $M_{{\rm H}_2}$ represents a lower limit due to the absorption or depletion of $^{12}$CO emission. 

To mitigate the effect of the foreground absorption, we employed the curve growth method on the redshifted component, which was less affected by the foreground absorption than the blueshifted one, to determine the flux and the radius of the gas disk, following \citet{Deng_2025}.
We used the integrated intensity map of Figure~\ref{fig:coline}(a) to estimate $F_{^{12}\rm CO,red}$ to be $351.1\pm5.9\,{\rm mJy\,km\,s^{-1}}$ and $R_{95\%,\rm gas}$ to be $98.4\pm10.5\,{\rm au}$, where the uncertainties were the RMS noise measured in the emission-free region and the standard deviation from the major axis of the CLEAN beam.
Assuming that $F_{^{12}\rm CO,red}$ is half of the total gas flux density, we then obtained the corrected total gas flux density of $F_{^{12}\rm CO}^{\rm corr}=702.2\pm5.9\,{\rm mJy\,km\,s^{-1}}$.
We also estimated the corrected gas mass $M_{\rm H_2}^{\rm corr}$, which was assumed to be twice as large as that of the redshifted one.
The corrected gas mass $M_{\rm H_2}^{\rm corr}$ was $2.1\times10^{-6}\,M_\odot$ in the LTE assumption and $5.8\times10^{-5}\,M_\odot$ in the case of the simple CO-to-H$_2$ conversion factor.
This result indicates that the estimate of gas mass may be affected by a factor of a few by foreground absorption. 
We therefore consider that the lower limit of gas mass is 0.2-5.8$\times 10^{-5}\,M_{\odot}$.
These values of the gas disk are summarized in Table~\ref{table:quantities}.
Our results imply that the inner region within the ring, where IRAS~04125+2902~b orbits, retains a non-negligible amount of gas.

\section{1D Brightness Distribution}\label{sec:disk_modeling}
\subsection{1D Visibility Analyses Methods}\label{subsec:vis_selection}
We constructed a 1D brightness model of the radial surface brightness distribution assuming that the intrinsic disk emission was axisymmetric.
We used \texttt{protomidpy}\footnote{\url{https://github.com/2ndmk2/protomidpy}}, \citep[][]{Aizawa_2024}, to reconstruct the radial surface brightness distribution that fits the observed visibility data.
The radial surface brightness profile is expanded using a Fourier-Bessel series whose coefficients are optimized to match the observed visibility.
This approach allows accurate modeling of the axisymmetric component while separating non-axisymmetric structures.
Furthermore, incorporating the Gaussian Process kernel helps estimate the smoothness of the brightness distribution, suppressing high-frequency noise and extracting the intrinsic features of the observational data.

The 1D brightness model requires geometric parameters, including central coordinates $\left(\Delta x_{\rm cen},\,\,\Delta y_{\rm cen}\right)$, an inclination angle $\cos i$, and a position angle ${\rm PA}$.
In addition, the Gaussian Process kernel has two hyperparameters $\alpha$ and $\gamma$ \citep[for details, see][]{Aizawa_2024}.
Among them, $\gamma$ is the spatial scale for regularization, which may be considered as a measure of spatial resolution in the reconstructed radial surface brightness profile.
These six parameters are simultaneously estimated using Bayesian inference with Markov Chain Monte Carlo (MCMC) to derive the optimal 1D model.

For the IRAS~04125+2902 1D brightness modeling, we used visibility data with baselines up to 1.2\,M$\lambda$ (see Figure~\ref{fig:model}a for the uv-caverage and Appendix~\ref{sec:data_flag}).
The 1D model image derived from the complete dataset showed artificial structures and strong side lobes, which hinder direct comparison with the continuum images in Figure~\ref{fig:continuum}.
To minimize the effect of artificial structures and sidelobes, we limited the baseline length up to 1.2\,M$\lambda$ (for details, see Appendix~\ref{sec:data_flag}).
For the MCMC approach, we set the number of iterations to 1000 and the number of walkers to 32. 
The initial values of the disk parameters were taken from $\cos i$ and PA measurements in \S\ref{subsec:dust}.

Figure~\ref{fig:model}(b) shows the image reconstructed based on the 1D brightness model.  
To investigate the features that are not recovered by the 1D brightness model, we subtracted the model visibility from the observed visibility constrained within the baseline of 1.2\,M$\lambda$.
Then, we applied the Fourier transformation using \texttt{tclean} with zero iterations (dirty map) to the subtracted visibility, and obtained the residual map shown in Figure~\ref{fig:model}(c).

Figure~\ref{fig:corner} presents the results of the MCMC model sampling.
We obtained the optimal parameters, $\left(\Delta x_{\rm cen},\,\,\Delta y_{\rm cen}\right)=\left(14.34^{+0.345}_{-0.385}, 3.64^{+0.415}_{-0.447}\right)$ in units of mas, $\cos i=0.825^{+0.0029}_{-0.0028}$, ${\rm PA}=174.93^{+0.472}_{-0.482}$\,deg, $\gamma=0.058^{+0.012}_{-0.014}$\,arcsec, and $\log_{10}\alpha=-3.42^{+0.158}_{-0.122}$.
The optimal values for $i$ and PA are comparable to those derived for $i_{\rm disk}$ and PA from the SpM image in \S\ref{subsec:dust}.

\begin{figure}[t]
    \centering
    \includegraphics[width=\linewidth]{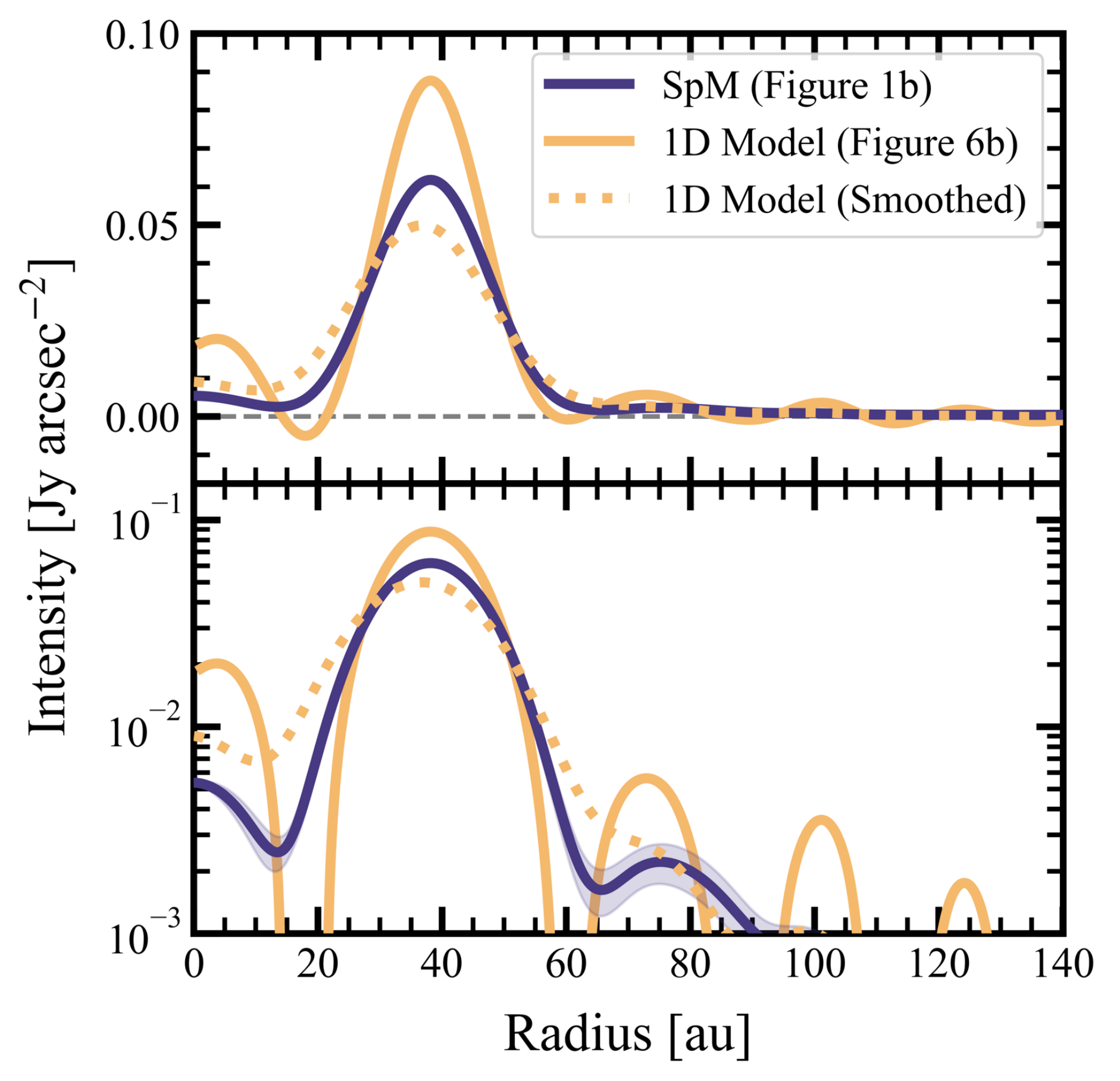}
    \caption{
    Same as the left panels of Figure~\ref{fig:cont_profile}, but for the radial intensity profile averaged over the full azimuthal angle of the 1D brightness model shown in Figure~\ref{fig:model}(b).
    The dotted yellow lines show the profiles of the 1D brightness model image smoothed with the effective spatial resolution $\theta_{\rm eff}$ by the CASA task \texttt{imsmooth}.
    }
    \label{fig:intensity}
\end{figure}

\subsection{Comparison with the Disk Model}\label{subsec:comparison}
We compare the SpM image (Figures~\ref{fig:continuum}b and c) with the 1D brightness model image (Figure~\ref{fig:model}b) to assess how well the model reproduces the structural features revealed by SpM imaging. 
Specifically, we examine the radial width and position of the outer ring, the presence and spatial extent of the inner emission within the gap, and any signs of non-axisymmetric structures.

\subsubsection{Outer Ring}\label{subsubsec:outer_ring}
Both the SpM and 1D brightness model images clearly exhibit a transitional disk with a ring-gap structure.
Figure~\ref{fig:intensity} shows the radial intensity profiles of the CLEAN, SpM, and 1D model images.
Using the same Gaussian fitting method described in \S\ref{subsec:dust} for the profile of the 1D model image, we estimated the ring width in the 1D model image to be 17.4$\pm$3.9\,au. 
The uncertainty reflected the standard deviation derived from the hyperparameter $\gamma$. 

The ring width measured in the SpM image (21.4$\pm$7.4\,au) differs from that of the 1D model image, primarily due to their difference in spatial resolution.
Note that 1D modeling and SpM are two independent methods of recovering spatial structures from observed visibility, so it is not surprising that their effective spatial resolutions are different.
As described by the hyperparameter $\gamma$ in \S~\ref{subsec:vis_selection}, the effective resolution of the 1D brightness model image (Figure~\ref{fig:model}b) may be about twice as high as that of the SpM image (Figures~\ref{fig:continuum}b and c).
To account for this, we convolved the 1D brightness model image to match the effective resolution of the SpM image. 
We found that the resulting radial intensity profile agrees reasonably well with that of the SpM image.
Similarly, the radial profile obtained from the SpM image matches with that of CLEAN after resolution matching (see Figure~\ref{fig:cont_profile}).
We confirmed a comparable trend, considering the RMS noise in the CLEAN image, further supporting the consistency between the two approaches after resolution matching (Figure~\ref{fig:cont_profile}).
This implies that the spatial resolution improves in the order of the CLEAN image, the SpM image, and the 1D brightness model image, and that even the SpM image could not fully resolve the outer ring.

\subsubsection{Inner Emission}\label{subsubsec:innter_emission}
The 1D brightness model image (Figure~\ref{fig:model}b) also reveals an inner emission component, consistent with the feature seen in the SpM reconstruction.  
As shown in Figure~\ref{fig:intensity}, the peak intensity of the inner emission is noticeably higher in the 1D brightness model image. 
This difference is likely due to its higher effective resolution ($\gamma$=0.058), which is approximately twice that of the SpM image.  
This also applies to the peak of the outer disk.
Despite the difference in the values of the intensity peak, both SpM and \texttt{protomidpy} consistently indicate the presence of weak emission within the gap.
We further discuss the origin of this emission in \S\ref{subsec:inner_disk}.

\subsubsection{Asymmetry}\label{subsubsec:asymmetry}
The axisymmetric disk modeling with \texttt{protomidpy} assumes an axisymmetric, geometrically thin disk and uses a one-dimensional Fourier–Bessel series to describe the radial intensity distribution. 
If the disk has any asymmetric structures, we may see them in the residual map shown in Figure~\ref{fig:model}(c).
However, it does not show any significant signal above 5.0$\times$10$^{-3}$\,mJy\,arcsec$^{-2}$ ($\sim$5$\sigma_{\rm dust}$). 
This suggests that the asymmetric features in the CLEAN and SpM images, if present at all, are very weak.

\section{Discussion}\label{sec:discussion}
\subsection{Presence of the Inner Disk}\label{subsec:inner_disk}
In the SpM and the disk model image of the IRAS~04125+2902 system, we could determine not only the clear ring-gap structure but also the weak dust emission within the gap. Figure~\ref{fig:intensity} shows that the emission within the gap has a peak intensity about ten times lower than that of the outer disk. 
The size of the weak emission region is comparable to the effective spatial resolution $\theta_{\rm eff}$ of the observation.
In this subsection, we discuss the origin of the weak emission within the ring, focusing on whether it is attributed to free-free emission or originates from the inner disk.

Free-free emission is a phenomenon commonly observed in the millimeter to centimeter wavelength range. 
\citet{Rota_2024} conducted a multiwavelength analysis of the continuum emission in the mm-cm wavelength range for 11 transitional disks. 
They found that in 10 cases, the central emission close to the star was consistent with free-free emission.

We checked previous cm-wavelength observations at 2-4\,GHz of IRAS~04125+2902 to investigate if the star exhibits free-free emission.
We extracted the dataset from the Very Large Array Sky Survey (VLASS) project \citep[][]{Lacy_2020}, which had an RMS noise level of 70\,$\mu$Jy\,beam$^{-1}$ and a spatial resolution of 2$\farcs$8$\times$2$\farcs$2 (448$\times$352\,au). 
As a result, no emission was found at all in the cm wavelength range.
With the most optimistic upper limit of 70\,$\mu$Jy (=RMS noise) for the inner emission at 2-4\,GHz in the cm wavelength range, the spectral index calculated from it and the flux of 0.13\,mJy measured in the SpM image was approximately 0.14. 
Therefore, we cannot rule out the possibility that the inner emission is due to free-free emission.  
Higher sensitivity observations at long wavelengths are needed.

\begin{figure}[t]
    \centering
    \includegraphics[width=\linewidth]{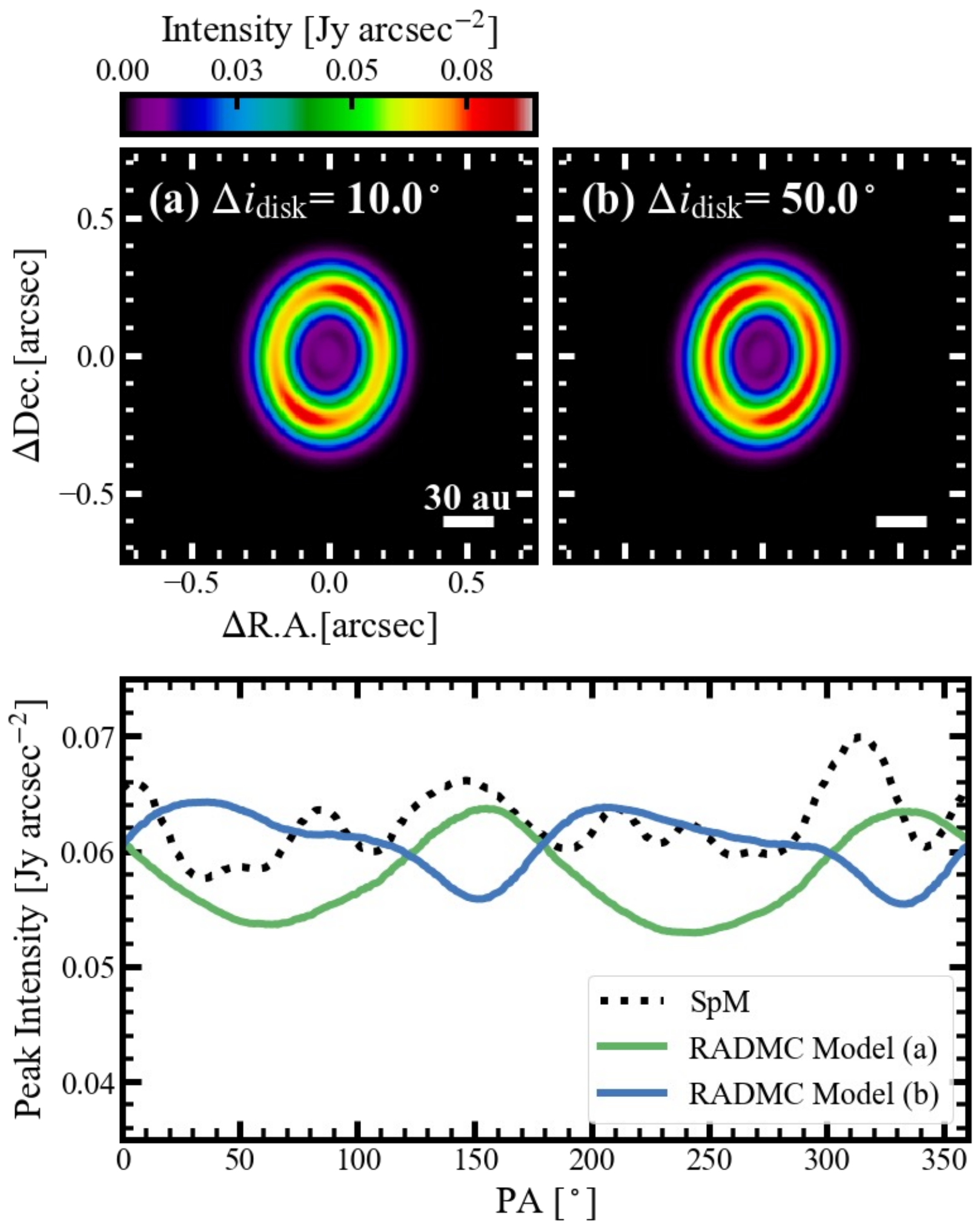}
    \caption{
    (Top) Two RADMC model images.
    Note that these images were convolved with a Gaussian beam of the same size as the effective spatial resolution $\theta_{\rm eff}$ of the SpM image.
    (Bottom) Peak intensity profile in the azimuthal direction.
    Note that all the profiles were created after the deprojection.
    All the profiles in both the top and bottom panels are linearly interpolated onto radial grid points spaced by 0.1\,au using \texttt{interpolate.interp1d} from the \texttt{SciPy} module.
    }
    \label{fig:radmc3d}
\end{figure}

\subsection{Inner Disk Shadow}\label{subsec:shadow}
The IRAS~04125+2902 harbors a transiting planet, which is misaligned with the ring. 
If the inner emission is due to an inner disk, the disk may not be aligned with the orbit of the planet, since the inner disk would block the star from view unless IRAS~04125+2902 is heavily reddened.
This motivates us to study whether the inner emission can instead be attributed to the inner disk, which may be misaligned with the outer disk.
If the inner disk is misaligned with the outer disk, it may partially obscure the stellar radiation.
This shadowing results in reduced heating in a portion of the ring, leading to a localized decrease in dust thermal emission.
The effect can explain the brightness distributions of some sources observed by ALMA or VLA \citep[e.g.,][]{Marino_2015,Orihara_2023}.

In the case of IRAS~04125+2902, we have only observed weak asymmetry in the image (Figure~\ref{fig:continuum}).
In addition, we could not confirm the existence of the asymmetric structure (Figure~\ref{fig:model}c) in the 1D brightness distribution presented in \S\ref{sec:disk_modeling}. 
Thus, we performed some radiative transfer simulations to estimate the degree of asymmetry that the inner disk introduces to the outer disk using \texttt{RADMC-3D} \citep[][]{Dullemond_2012}.
In this paper, we only consider the effects of different inclination angles between the inner and the outer disk for demonstrative purposes.

In this calculation, we used polar coordinates with each grid divided into 128 segments. 
For the physical disk structure, we adopted the dust temperature profile $T_{\rm dust}$ as,
\begin{align}
    T_{\rm dust}=T_0\times\left(\frac{r}{1\,{\rm au}}\right)^{-0.5},\label{eq:temp_profile}
\end{align}
where $r$ is the distance from the central star, and $T_0 = 230$\,K is the temperature at $r$=1.0\,au, estimated from the stellar luminosity $L_\ast$=0.4-0.5\,$L_\odot$ \citep[][]{Espaillat_2015, Testi_2022}.

Next, the scale height $H$ was calculated as,
\begin{align}
    H=\frac{1}{f_{\rm set}}\frac{c_s}{\Omega}=\frac{1}{f_{\rm set}}\times\sqrt{\frac{k T_{\rm dust}}{\mu m_{\rm H}}}\times\sqrt{\frac{r^3}{GM_\ast}},\label{eq:scale_height}
\end{align}
where $f_{\rm set}$ is the dust settling factor and we set $f_{\rm set}$=1.0 for simplicity purposes.
In Equation (\ref{eq:scale_height}), $c_s$ is the sound speed, $\Omega$ is the Keplarian angular velocity, $\mu$ is the average molecular weight ($\mu$=2.34), $m_{\rm H}$ is the mass of a hydrogen atom, $k$ is the Boltzman constant, $G$ is the gravitational constant, and $M_\ast$ is the stellar mass ($M_\ast$=1.0\,$M_\odot$, see \S\ref{subsec:12co}).

Then, the surface densities of the inner and outer disks, $\Sigma_{\rm d, in}$ and $\Sigma_{\rm d, out}$, were assumed as, 
\begin{align}
    \Sigma_{\rm d, in}&=\Sigma_{\rm peak, in}\exp{\left[-\frac{\left(r-r_{\rm peak, in}\right)^2}{2\sigma_{\rm in}^2}\right]},\label{eq:sden_in}\\
    \Sigma_{\rm d, out}&=\Sigma_{\rm peak, out}\exp{\left[-\frac{\left(r-r_{\rm peak, out}\right)^2}{2\sigma_{\rm out}^2}\right]}.\label{eq:sden_out}
\end{align}
In Equation~(\ref{eq:sden_in}), $\Sigma_{\rm peak, in}$, $r_{\rm preak, in}$, and $\sigma_{\rm in}$ represent the peak surface density, its position, and the width of the inner disk, respectively.
We used the fixed values of $r_{\rm preak, in}$=3.0\,au and $\sigma_{\rm in}$=2.0\,au.
$\Sigma_{\rm peak, out}$, $r_{\rm peak, out}$, and $\sigma_{\rm out}$ in Equation~(\ref{eq:sden_out}) correspond to the parameters of the outer disk with a ring structure, where we also adopted $r_{\rm peak, outer}$=41.0\,au and $\sigma_{\rm outer}$=5.0\,au in our calculation.
\begin{table}[t]
    \centering
    \caption{Parameters of radiative transfer models}
    \label{table:model_para}
    \begin{tabular}{ccccc}
    \hline
      & $\Sigma_{\rm peak, in}$ & $\Sigma_{\rm peak, out}$ & $\Delta i_{\rm disk}$ & $\Delta$PA \\
      & $\left[{\rm g\,cm^{-2}}\right]$ & $\left[{\rm g\,cm^{-2}}\right]$ & $\left[^\circ\right]$ & $\left[^\circ\right]$  \\
    \hline
    \hline
     RADMC Model (a) & 0.09 & 1.70 & 10.0 & -18.8\\
     RADMC Model (b) & 0.13 & 1.55 & 50.0 & -18.8\\
     \hline
     \hline
    \end{tabular}
\end{table}

For the simulation of the misalignment between the inner and outer disks, we applied parameters $\Delta i_{\rm disk}$ and $\Delta$PA, which represent the differences in inclination angles and PA of the inner disk relative to those of the outer disk.
We assumed $i_{\rm in}=i_{\rm disk}+\Delta i_{\rm disk}$ and ${\rm PA}_{\rm in}={\rm PA}_{\rm disk}+\Delta{\rm PA}$, where $i_{\rm in}$ and ${\rm PA}_{\rm in}$ are the inclination angle and PA of the inner disk, and $i_{\rm disk}$ and ${\rm PA}_{\rm disk}$ are those of the outer disk, as estimated in \S\ref{subsec:dust}.
For demonstrative purposes, we fixed $\Delta$PA to be -18.8$^{\circ}$ and varied $\Delta i_{\rm disk}$ only to see qualitative trends.
To estimate $\Delta$PA, we assume that the weak dip in the SpM image seen at PA=156.3$^\circ$ from the right panel of Figure~\ref{fig:cont_profile}.
We assumed that the major axis of the inner disk was consistent with the direction of the dip, which was offset by -18.8$^\circ$ from the major axis of the outer ring.
For $\Delta i_{\rm disk}$, we presented the cases with 10$^{\circ}$ and 50$^{\circ}$.
The case of $\Delta i_{\rm disk}$=50.0$^\circ$ corresponds to the inner disk with an inclination angle of 85.6$^\circ$, i.e., the inner disk is nearly edge-on.
The flux densities in the models vary with different values of $\Delta i_{\rm disk}$.
With $\Sigma_{\rm d, out}$ fixed at 1.5-1.7\,g\,cm$^{-2}$, we adjusted $\Sigma_{\rm d, in}$ so that the final model image reproduces the flux densities within $r<$ 15\,au and $r<$ 60\,au measured in the SpM image of Figure~\ref{fig:channel}(b), which are approximately 0.13\,mJy and 13.0\,mJy, respectively.
Table~\ref{table:model_para} summarizes the final parameters for the radiative transfer models.
For dust opacity, we used the models of amorphous olivine with 50\% Mg and 50\% Fe \citep[][]{Jaeger_1994, Dorschner_1995}, provided in \texttt{RADMC-3D} package.

We used \texttt{RADMC-3D} \verb|mctherm| run to recalculate the temperature profile of the disk and then obtained a model image. 
We set the values of the inclination and position angle of the system to be the same as those of the outer disk in the SpM image.
After the calculation, we convolved the model image with a Gaussian beam of the same size as the effective spatial resolution $\theta_{\rm eff}$ of the SpM image using \texttt{convolution.Gaussian2DKernel} and \texttt{convolution.convolve} from the \texttt{astropy} module.

Figure~\ref{fig:radmc3d} shows two RADMC model images and their intensity profiles from the radiative transfer calculations.
The results of radiative transfer calculations show that both cases in Figure~\ref{fig:radmc3d}, with small or large misalignment in the inclination angle, exhibit weak asymmetry.  
The level of asymmetry is comparable with that in the observed images (SpM or CLEAN).
The peak intensity profile of the RADMC model seems to match better in the case of $\Delta i_{\rm disk}$=10$^\circ$ than in the case of $\Delta i_{\rm disk}$=50$^\circ$, while both models do not match perfectly with the observations.
Nevertheless, since the inner disk is not fully spatially resolved in the SpM image and the physical properties of the inner disk remain uncertain, we note that this analysis is preliminary.
To evaluate the more realistic physical properties of the inner disk, high sensitivity and resolution observations are required to constrain the physical parameters more accurately.
Higher spatial resolution and sensitivity, as well as creating and comparing intensity profiles, will be important.

The small misalignment of the inner disk relative to the outer disk is consistent with the results of $^{12}$CO $J$=2--1 shown in Figures~\ref{fig:channel} and \ref{fig:coline}.
In the IRAS~04125+2902 system, the $^{12}$CO $J$=2--1 velocity field in Figures~\ref{fig:channel} and \ref{fig:coline}(b) does not exhibit any significant distortion near the central star.
If the difference in inclination angles between the inner and outer disks is small, their Keplerian rotations should be nearly identical, meaning that there should be no significant distortion in the velocity structure.
The misalignment of 20-40$^{\circ}$ between the inner and the outer disk would lead to the velocity distortion of a detectable level \citep[see e.g.,][for the case of SY~Cha]{Orihara_2023}.
Therefore, in comparison with SY~Cha, it is not surprising that such distortion is not observed in the IRAS~04125+2902 system.
However, since $^{12}$CO $J$=2--1 emission around IRAS~04125+2902 is likely affected by foreground absorption, it is necessary to use other lines for better investigations of gas kinematics.

Finally, we examine the optical depth and dust mass in our RADMC model.  
The optical depth at 1.3\,mm in the model of $\Delta i_{\rm disk}=10\,^\circ$ is below 0.25 even at its peak, suggesting that the disk is globally optically thin.
This indicates that the optically thin assumption of Equation~(\ref{eq:dust_mass}) in \S\ref{subsec:dust} is reasonable.
However, the input dust mass of the RADMC model is $6.2 \times 10^{-4}~M_{\odot}$, which is $\sim 20$ times larger than the dust mass estimate in \S\ref{subsec:dust}.
This is because the dust absorption opacity at 1.3\,mm in the RADMC model is 0.12\,cm$^{2}$\,g$^{-1}$ (and the scattering opacity is negligibly small) while that used in \S\ref{subsec:dust} is 2.3\,cm$^{2}$\,g$^{-1}$.  
This factor of 20 opacity difference explains the difference in the dust mass in \S\ref{subsec:dust} and the RADMC model.  
We need multi-wavelength data to break the degeneracy of disk mass, opacity, and temperature.

\subsection{Implications for the Formation of the Misaligned System}\label{subsec:misalignment}
The IRAS~04125+2902 system is a highly complex misaligned system, in which the orbital axes of the binary companion, the outer disk, the inner disk, and the transiting planet are all mutually misaligned.
\citet{Barber_2024} proposed that such a configuration may result from gravitational scattering by an additional massive planet in the system, which was initially aligned with the disk and the transiting planet \citep[for details, see][]{Nagasawa_2008}. 
However, no such outer massive planet has been detected to date, and our results do not show any evidence supporting its presence.

An alternative and promising scenario, which was also pointed out by \citet{Barber_2024}, is that the disk became warped through interactions between the early disk and the surrounding environment \citep[e.g,][]{Bate_2018}. 
In such a case, the inclination angle of the disk could vary with radius, and planets forming within the disk might naturally acquire different orbital inclinations depending on their formation locations.
Recently, \citet{Hirano_2020} and \citet{Machida_2020} have indicated that many disks should be warped when considering misalignment between the rotational axis and magnetic field in prestellar clouds.

New high-resolution ALMA observations have enabled a precise measurement of the inclination angle of the outer disk and have revealed the likely presence of a misaligned inner disk. 
In addition, our radiative transfer modeling reproduces a configuration in which the inner and outer disks may not be fully aligned (see \S\ref{subsec:shadow}), providing observational support for the warped disk scenario proposed by \citet{Barber_2024}.
Furthermore, the derived (lower limit of ) gas mass might be comparable to or even lower than the dust mass, suggesting that the disk is in a highly evolved state (see Table~\ref{table:quantities}). 
If more gas had been present during the planet formation phase, the formation of the warped disk may be more plausible.

Moreover, \citet{Nealon_2025} proposed an alternative explanation for the complex misaligned system of IRAS~04125+2902, involving the encounter with a low-mass star and either a three-body interaction or von Zeipel–Kozai–Lidov oscillations.
Our measurement of gas disk size ($\sim$100\,au) indicates that three-body interaction may play a role in shaping the system in their scenario.
The prediction of the stellar encounter scenario is the misalignment between the binary and the planet's orbit.  
In our study, we detected inner emission that could be attributed to an inner disk that is not strongly misaligned with the outer ring (see \S\ref{subsec:inner_disk} and \ref{subsec:shadow}).
This implies that even in the region as close as $\lesssim$15\,au from the central star (i.e., effective spatial resolution of the SpM image), the disk does not show a clear tendency for alignment with the binary orbit.
Therefore, we consider that the stellar encounter scenario is not ruled out by our observations, at least qualitatively.
Predictions for the inner disk geometry may be required for further comparison between the scenario and observations.
Therefore, at present, it is not possible to prove that the binary and planet orbits are aligned, and the scenario proposed by \citet{Nealon_2025} should be regarded as one possible explanation for the misaligned system.

In summary, while not yet conclusive, our observations are consistent with the second scenario proposed by \citet{Barber_2024} and with the scenario proposed by \citet{Nealon_2025}.
Our observations have put new constraints on the outer disk geometry and indicated the presence of an inner disk that is probably more aligned with the outer disk than the transiting planet.
Higher angular resolution and sensitivity observations are needed to put more constraints, especially on the properties of the emission close to the central star.

\renewcommand{\labelenumi}{\arabic{enumi})}
\section{Summary}\label{sec:summary}
We used high-resolution archival ALMA Band 6 data to study the protoplanetary disk around IRAS~04125+2902, the youngest system known to host a transiting planet.

Our main findings are summarized as follows:
\begin{enumerate}
   \item
   The analysis of the dust continuum emission revealed the transitional disk with a clear ring-gap structure. 
   Furthermore, in the image generated by super-resolution imaging with SpM, we confirmed weak emission within the gap and a weak asymmetric structure on the northwest side of the ring.
   We performed symmetric disk modeling with \texttt{protomidpy} for the dust continuum dataset, and the result supported the weak dust emission within the gap shown in the SpM image. 
   \item
   The inclination angle of the outer ring was measured to be 35.6$\pm$0.2$^\circ$, indicating that the disk is nearly face-on and misaligned with the proper motions of the binary companion and the orbital motion of the transiting planet with inclination angles of $\sim$90$^\circ$. 
   From the flux density, the lower limit of the dust mass for the transitional disk was estimated to be $\sim$3.0$\times$10$^{-5}$\,$M_\odot$ ($\sim$10.1$M_\oplus$).
   \item
   We detected $^{12}$CO $J$=2--1 emission around IRAS~04125+2902 in the velocity range of 4.8 to 12.6\,km\,s$^{-1}$. 
   The blueshifted and redshifted components trace the Keplerian rotation of the gas disk with a stellar mass of 0.7--1.0\,$M_\odot$. 
   The gas mass was estimated to be $\sim$0.2--5.8$\times$10$^{-5}$\,$M_\odot$, which should be considered as a lower limit. 
   \item
   A previous VLA observation with a spatial resolution of $\sim$2$\farcs$5 did not show any emission; however, we cannot completely rule out that the weak emission was caused by free-free emission.
   We require further observations with high sensitivity at cm wavelengths for detailed verification.
   \item
   Assuming the existence of the inner disk misaligned with the outer disk, we performed radiative transfer calculations for the IRAS~04125+2902 system. 
   The results showed that a small misalignment in the inclination angle of $\sim$10$^\circ$ with respect to the outer disk may explain the weak asymmetric structure shown in Figure~\ref{fig:continuum}.
   The small misalignment is consistent with the fact that the velocity field of $^{12}$CO $J$=2--1 emission does not exhibit any significant disturbances near the central star.
   Further observations for the dust continuum emission and other lines with higher sensitivity and spatial resolution are necessary to investigate the asymmetry and gas kinematics in detail.
\end{enumerate}

The IRAS~04125+2902 likely exhibits a dynamically complex structure.  The binary orbit, the disk, and the transiting planets are misaligned.
There also may be an internal structure within the disk, i.e., the outer ring that is possibly misaligned with the inner disk.
Such a multilayered misalignment may influence planet formation processes, suggesting a greater diversity in the environments where planets can form.
Furthermore, since IRAS~04125+2902 hosts one of the youngest transiting planets known to date, our study provides valuable information on the orbital evolution of a young protoplanet.

\begin{acknowledgments}
The authors appreciate the anonymous referee for all of the comments and advice that helped improve the manuscript and the contents of this study.
The authors thank Dr. Masataka Aizawa for his technical support and Dr. Masayuki Yamaguchi for valuable discussion.
This work was supported by a NAOJ ALMA Scientific Research grant (No. 2022-22B; MNM) and by JSPS KAKENHI 25KJ1947 (AS), 23K03463 (TM), 25KJ1921 (MO), 20H05645 (KT), 21H00049 (KT), and 21K13962 (KT).
G.D.M. acknowledges support from FONDECYT project 1252141 and the ANID BASAL project FB210003.
This paper makes use of the following ALMA data: ADS/JAO.ALMA\#2022.1.01302.S ALMA is a partnership of ESO (representing its member states), NSF (USA) and NINS (Japan), together with NRC (Canada), MOST and ASIAA (Taiwan), and KASI (Republic of Korea), in cooperation with the Republic of Chile. 
The Joint ALMA Observatory is operated by ESO, AUI/NRAO and NAOJ.
The National Radio Astronomy Observatory is a facility of the National Science Foundation operated under cooperative agreement by Associated Universities, Inc.
Data analysis was in part carried out on a common-use data analysis computer system at the Astronomy Data Center, ADC, of the National Astronomical Observatory of Japan.
\end{acknowledgments}

\facility{ALMA, VLA}

\software{astropy \citep[e.g.,][]{astropy_2022},
          CASA \citep[][]{CASA_2022},
          matplotlib \citep[][]{Hunter_2007},
          PRIISM \citep[][]{Nakazato_2020,Nakazato_2020b},
          protomidpy \citep[][]{Aizawa_2024},
          SciPy \citep[][]{Virtanen_2020},
          RADMC-3D \citep[][]{Dullemond_2012},
          }

\setcounter{figure}{0}
\renewcommand{\thefigure}{A.\arabic{figure}}
\begin{figure*}
    \centering
    \includegraphics[width=\linewidth]{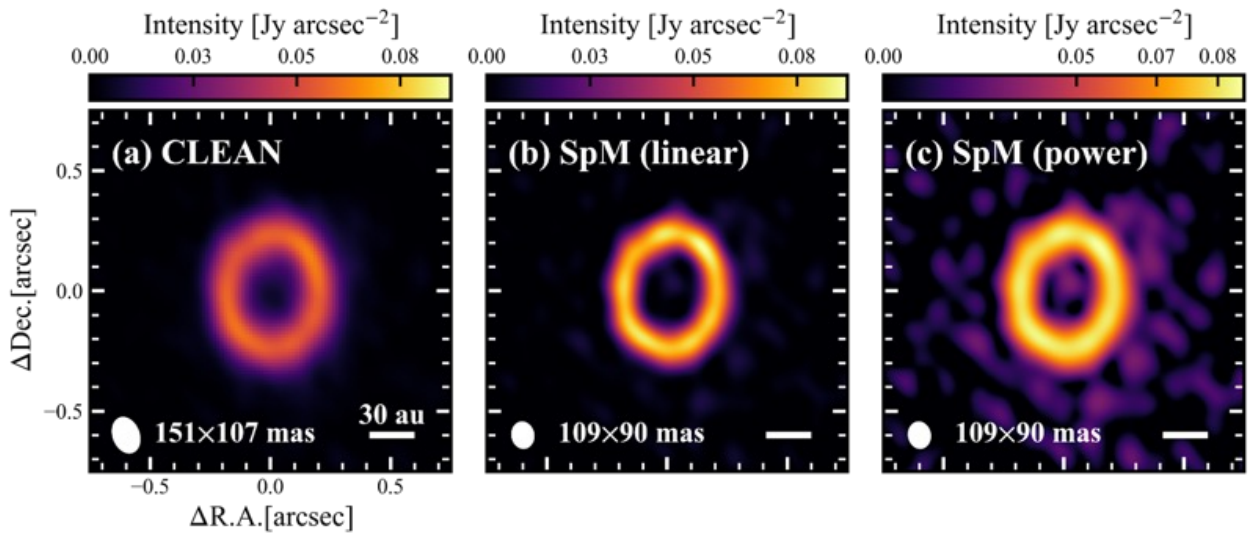}
    \caption{Same as Figure~\ref{fig:continuum}, but we adopted the same colorscale as the SpM image for the CLEAN image.}
    \label{fig:cont_app}
\end{figure*}

\setcounter{figure}{0}
\renewcommand{\thefigure}{B.\arabic{figure}}
\begin{figure*}[ht]
    \centering
    \includegraphics[width=\linewidth]{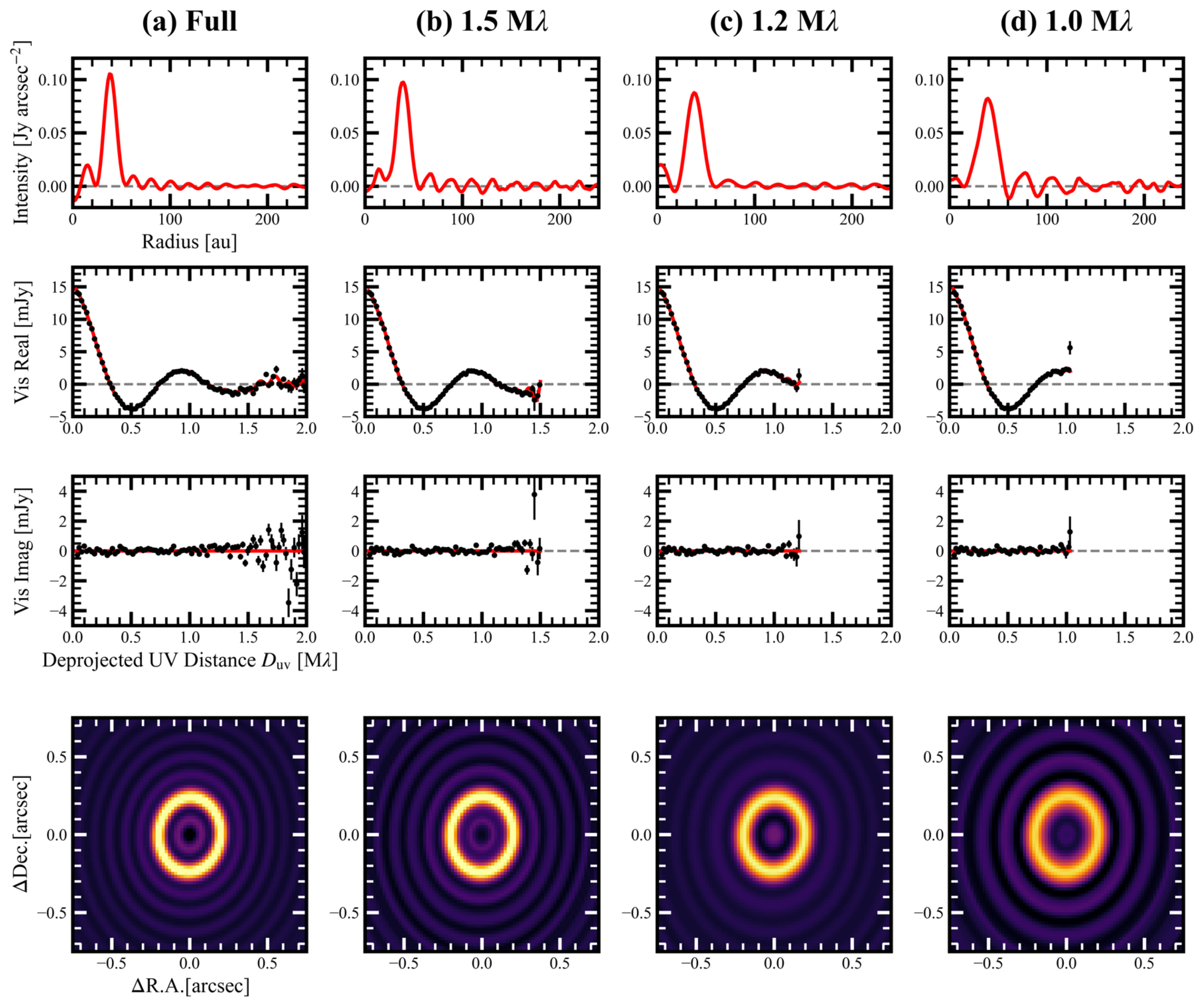}
    \caption{
    Results of the disk modeling based on \citet{Aizawa_2024} applied to four different datasets: full-baseline observation visibility (1st column), and visibility limited to baselines of 1.5\,M$\lambda$ (2nd column), 1.2\,M$\lambda$ (3rd column), and 1.0\,M$\lambda$ (4th column).
    From top to bottom, the panels show the intensity profiles of the best-fit disk model images, the real and imaginary parts of the observation and model visibilities, and two-dimensional images created from the intensity profiles assuming an axisymmetric disk displayed with the same color scale.
    In the visibility plots, black markers indicate the observed visibilities binned into 107 bins with a width of 20\,k$\lambda$ from 0.1\,$\lambda$ to 2.3\,M$\lambda$, while red lines represent the model visibilities.
    }
    \label{fig:comparison}
\end{figure*}

\appendix
\section{Detailed Comparison of 1.3\,mm Dust Continuum images}\label{sec:dust_comp}
Figure~\ref{fig:cont_app}(a) shows the CLEAN image of Figure~\ref{fig:continuum}(a) on the same color scale as the SpM image shown in Figures~\ref{fig:continuum}(b) and \ref{fig:cont_app}(b).
The CLEAN image of Figure~\ref{fig:cont_app}(a) appears to have a lower brightness distribution than the SpM image.
This is because the spatial resolution of the CLEAN image is lower than that of the SpM image, so the beam dilution effects are more significant for the ring structure that is not fully spatially resolved.

\section{Data Flag for Disk Modeling}\label{sec:data_flag}
The panels in the first column of Figure~\ref{fig:comparison} show the results after applying disk modeling to the full dataset.
The disk model closely matches the real part of the observed visibility. 
The two-dimensional model image also reveals several sidelobes and inner and outer ring structures.
However, the intensity profile exhibits significant negative components inside the inner ring.
Figure~\ref{fig:model}(a) illustrates that the distribution of the observed visibility is concentrated in a particular direction.
When part of the $uv$ plane is truncated to create an image, the missing visibilities are assumed to be zero, and this discontinuity can produce high-frequency oscillations such as sidelobes in the Fourier-transformed image.
As a result, we confirmed that an artificial structure, which did not physically exist, was reproduced, along with a negative brightness distribution, to conserve the total flux.
Thus, we did not adopt the disk model derived from the full observed visibility.

We performed disk modeling by limiting the baseline length to 1.5\,M$\lambda$, 1.2\,M$\lambda$, and 2.0\,M$\lambda$ and determined the appropriate disk model that minimizes sidelobe effects and avoids artificial structures.
Figure~\ref{fig:comparison} shows the results in the right three columns for baseline lengths of 1.5\,M$\lambda$, 1.2\,M$\lambda$, and 1.0\,M$\lambda$.
The results for the dataset with a 1.5\,M$\lambda$ baseline show a similar profile to that of the full dataset, though with stronger sidelobes.
In the model with a 1.0\,M$\lambda$ baseline, there is no inner emission, and no negative intensity is present in the region of small radii. 
However, the sidelobes exhibit strong intensity and negative components, making them unsuitable for comparison with the SpM image.
In contrast, the model with a 1.2\,M$\lambda$ baseline exhibits the fewest negative components and sidelobes with relatively weak intensity. 
This is because the visibility with a 1.2\,M$\lambda$ baseline, shown by the red circle in Figure~\ref{fig:model}(a), is isotropically distributed. 
The distribution minimizes the influence of $uv$ sampling.
At the same time, it conserves the high-frequency components.
Therefore, we adopted the model derived from the observation visibility with a baseline limited to 1.2\,M$\lambda$ in \S\ref{subsec:vis_selection}.

\bibliography{reference}{}

\begin{thebibliography}{}
\expandafter\ifx\csname natexlab\endcsname\relax\def\natexlab#1{#1}\fi
\providecommand{\url}[1]{\href{#1}{#1}}
\providecommand{\dodoi}[1]{doi:~\href{http://doi.org/#1}{\nolinkurl{#1}}}
\providecommand{\doeprint}[1]{\href{http://ascl.net/#1}{\nolinkurl{http://ascl.net/#1}}}
\providecommand{\doarXiv}[1]{\href{https://arxiv.org/abs/#1}{\nolinkurl{https://arxiv.org/abs/#1}}}

\bibitem[{{Aizawa} {et~al.}(2024){Aizawa}, {Muto}, \& {Momose}}]{Aizawa_2024}
{Aizawa}, M., {Muto}, T., \& {Momose}, M. 2024, \mnras, 532, 1361, \dodoi{10.1093/mnras/stae1549}

\bibitem[{{Akiyama} {et~al.}(2017){Akiyama}, {Kuramochi}, {Ikeda}, {Fish}, {Tazaki}, {Honma}, {Doeleman}, {Broderick}, {Dexter}, {Mo{\'s}cibrodzka}, {Bouman}, {Chael}, \& {Zaizen}}]{Akiyama_2017}
{Akiyama}, K., {Kuramochi}, K., {Ikeda}, S., {et~al.} 2017, \apj, 838, 1, \dodoi{10.3847/1538-4357/aa6305}

\bibitem[{{Ansdell} {et~al.}(2016){Ansdell}, {Williams}, {van der Marel}, {Carpenter}, {Guidi}, {Hogerheijde}, {Mathews}, {Manara}, {Miotello}, {Natta}, {Oliveira}, {Tazzari}, {Testi}, {van Dishoeck}, \& {van Terwisga}}]{Ansdell_2016}
{Ansdell}, M., {Williams}, J.~P., {van der Marel}, N., {et~al.} 2016, \apj, 828, 46, \dodoi{10.3847/0004-637X/828/1/46}

\bibitem[{{Astropy Collaboration} {et~al.}(2022){Astropy Collaboration}, {Price-Whelan}, {Lim}, {Earl}, {Starkman}, {Bradley}, {Shupe}, {Patil}, {Corrales}, {Brasseur}, {N{"o}the}, {Donath}, {Tollerud}, {Morris}, {Ginsburg}, {Vaher}, {Weaver}, {Tocknell}, {Jamieson}, {van Kerkwijk}, {Robitaille}, {Merry}, {Bachetti}, {G{"u}nther}, {Aldcroft}, {Alvarado-Montes}, {Archibald}, {B{'o}di}, {Bapat}, {Barentsen}, {Baz{'a}n}, {Biswas}, {Boquien}, {Burke}, {Cara}, {Cara}, {Conroy}, {Conseil}, {Craig}, {Cross}, {Cruz}, {D'Eugenio}, {Dencheva}, {Devillepoix}, {Dietrich}, {Eigenbrot}, {Erben}, {Ferreira}, {Foreman-Mackey}, {Fox}, {Freij}, {Garg}, {Geda}, {Glattly}, {Gondhalekar}, {Gordon}, {Grant}, {Greenfield}, {Groener}, {Guest}, {Gurovich}, {Handberg}, {Hart}, {Hatfield-Dodds}, {Homeier}, {Hosseinzadeh}, {Jenness}, {Jones}, {Joseph}, {Kalmbach}, {Karamehmetoglu}, {Ka{l}uszy{'n}ski}, {Kelley}, {Kern}, {Kerzendorf}, {Koch}, {Kulumani}, {Lee}, {Ly}, {Ma}, {MacBride}, {Maljaars}, {Muna}, {Murphy}, {Norman}, {O'Steen},
  {Oman}, {Pacifici}, {Pascual}, {Pascual-Granado}, {Patil}, {Perren}, {Pickering}, {Rastogi}, {Roulston}, {Ryan}, {Rykoff}, {Sabater}, {Sakurikar}, {Salgado}, {Sanghi}, {Saunders}, {Savchenko}, {Schwardt}, {Seifert-Eckert}, {Shih}, {Jain}, {Shukla}, {Sick}, {Simpson}, {Singanamalla}, {Singer}, {Singhal}, {Sinha}, {Sip{H{o}}cz}, {Spitler}, {Stansby}, {Streicher}, {{{S}}umak}, {Swinbank}, {Taranu}, {Tewary}, {Tremblay}, {Val-Borro}, {Van Kooten}, {Vasovi{'c}}, {Verma}, {de Miranda Cardoso}, {Williams}, {Wilson}, {Winkel}, {Wood-Vasey}, {Xue}, {Yoachim}, {Zhang}, {Zonca}, \& {Astropy Project Contributors}}]{astropy_2022}
{Astropy Collaboration}, {Price-Whelan}, A.~M., {Lim}, P.~L., {et~al.} 2022, \apj, 935, 167, \dodoi{10.3847/1538-4357/ac7c74}

\bibitem[{{Barber} {et~al.}(2024){Barber}, {Mann}, {Vanderburg}, {Krolikowski}, {Kraus}, {Ansdell}, {Pearce}, {Mace}, {Andrews}, {Boyle}, {Collins}, {De Furio}, {Dragomir}, {Espaillat}, {Feinstein}, {Fields}, {Jaffe}, {Lopez Murillo}, {Murgas}, {Newton}, {Palle}, {Sawczynec}, {Schwarz}, {Thao}, {Tofflemire}, {Watkins}, {Jenkins}, {Latham}, {Ricker}, {Seager}, {Vanderspek}, {Winn}, {Charbonneau}, {Essack}, {Rodriguez}, {Shporer}, {Twicken}, \& {Villase{\~n}or}}]{Barber_2024}
{Barber}, M.~G., {Mann}, A.~W., {Vanderburg}, A., {et~al.} 2024, \nat, 635, 574, \dodoi{10.1038/s41586-024-08123-3}

\bibitem[{{Bate}(2018)}]{Bate_2018}
{Bate}, M.~R. 2018, \mnras, 475, 5618, \dodoi{10.1093/mnras/sty169}

\bibitem[{{Beckwith} {et~al.}(1990){Beckwith}, {Sargent}, {Chini}, \& {Guesten}}]{Beckwith_1990}
{Beckwith}, S. V.~W., {Sargent}, A.~I., {Chini}, R.~S., \& {Guesten}, R. 1990, \aj, 99, 924, \dodoi{10.1086/115385}

\bibitem[{{Benisty} {et~al.}(2021){Benisty}, {Bae}, {Facchini}, {Keppler}, {Teague}, {Isella}, {Kurtovic}, {P{\'e}rez}, {Sierra}, {Andrews}, {Carpenter}, {Czekala}, {Dominik}, {Henning}, {Menard}, {Pinilla}, \& {Zurlo}}]{Benisty_2021}
{Benisty}, M., {Bae}, J., {Facchini}, S., {et~al.} 2021, \apjl, 916, L2, \dodoi{10.3847/2041-8213/ac0f83}

\bibitem[{{Bolatto} {et~al.}(2013){Bolatto}, {Wolfire}, \& {Leroy}}]{Balatto_2013}
{Bolatto}, A.~D., {Wolfire}, M., \& {Leroy}, A.~K. 2013, \araa, 51, 207, \dodoi{10.1146/annurev-astro-082812-140944}

\bibitem[{{CASA Team} {et~al.}(2022){CASA Team}, {Bean}, {Bhatnagar}, {Castro}, {Donovan Meyer}, {Emonts}, {Garcia}, {Garwood}, {Golap}, {Gonzalez Villalba}, {Harris}, {Hayashi}, {Hoskins}, {Hsieh}, {Jagannathan}, {Kawasaki}, {Keimpema}, {Kettenis}, {Lopez}, {Marvil}, {Masters}, {McNichols}, {Mehringer}, {Miel}, {Moellenbrock}, {Montesino}, {Nakazato}, {Ott}, {Petry}, {Pokorny}, {Raba}, {Rau}, {Schiebel}, {Schweighart}, {Sekhar}, {Shimada}, {Small}, {Steeb}, {Sugimoto}, {Suoranta}, {Tsutsumi}, {van Bemmel}, {Verkouter}, {Wells}, {Xiong}, {Szomoru}, {Griffith}, {Glendenning}, \& {Kern}}]{CASA_2022}
{CASA Team}, {Bean}, B., {Bhatnagar}, S., {et~al.} 2022, \pasp, 134, 114501, \dodoi{10.1088/1538-3873/ac9642}

\bibitem[{{Currie} {et~al.}(2022){Currie}, {Lawson}, {Schneider}, {Lyra}, {Wisniewski}, {Grady}, {Guyon}, {Tamura}, {Kotani}, {Kawahara}, {Brandt}, {Uyama}, {Muto}, {Dong}, {Kudo}, {Hashimoto}, {Fukagawa}, {Wagner}, {Lozi}, {Chilcote}, {Tobin}, {Groff}, {Ward-Duong}, {Januszewski}, {Norris}, {Tuthill}, {van der Marel}, {Sitko}, {Deo}, {Vievard}, {Jovanovic}, {Martinache}, \& {Skaf}}]{Currie_2022}
{Currie}, T., {Lawson}, K., {Schneider}, G., {et~al.} 2022, Nature Astronomy, 6, 751, \dodoi{10.1038/s41550-022-01634-x}

\bibitem[{{Deng} {et~al.}(2025){Deng}, {Vioque}, {Pascucci}, {P{\'e}rez}, {Zhang}, {Kurtovic}, {Trapman}, {TorresVillanueva}, {Agurto-Gangas}, {Carpenter}, {Pinilla}, {Gorti}, {Tabone}, {Sierra}, {Rosotti}, {Cieza}, {Anania}, {Gonz{\'a}lez-Ruilova}, {Hogerheijde}, {Miley}, {Ruiz-Rodriguez}, {Ruaud}, \& {Schwarz}}]{Deng_2025}
{Deng}, D., {Vioque}, M., {Pascucci}, I., {et~al.} 2025, arXiv e-prints, arXiv:2506.10734, \dodoi{10.48550/arXiv.2506.10734}

\bibitem[{{Dorschner} {et~al.}(1995){Dorschner}, {Begemann}, {Henning}, {Jaeger}, \& {Mutschke}}]{Dorschner_1995}
{Dorschner}, J., {Begemann}, B., {Henning}, T., {Jaeger}, C., \& {Mutschke}, H. 1995, \aap, 300, 503

\bibitem[{{Dullemond} {et~al.}(2012){Dullemond}, {Juhasz}, {Pohl}, {Sereshti}, {Shetty}, {Peters}, {Commercon}, \& {Flock}}]{Dullemond_2012}
{Dullemond}, C.~P., {Juhasz}, A., {Pohl}, A., {et~al.} 2012, {RADMC-3D: A multi-purpose radiative transfer tool}, Astrophysics Source Code Library, record ascl:1202.015

\bibitem[{{Dunham} {et~al.}(2014){Dunham}, {Arce}, {Mardones}, {Lee}, {Matthews}, {Stutz}, \& {Williams}}]{Dunham_2014}
{Dunham}, M.~M., {Arce}, H.~G., {Mardones}, D., {et~al.} 2014, \apj, 783, 29, \dodoi{10.1088/0004-637X/783/1/29}

\bibitem[{{Espaillat} {et~al.}(2015){Espaillat}, {Andrews}, {Powell}, {Feldman}, {Qi}, {Wilner}, \& {D'Alessio}}]{Espaillat_2015}
{Espaillat}, C., {Andrews}, S., {Powell}, D., {et~al.} 2015, \apj, 807, 156, \dodoi{10.1088/0004-637X/807/2/156}

\bibitem[{{Evans} {et~al.}(2022){Evans}, {Kim}, \& {Ostriker}}]{Evans_2022}
{Evans}, N.~J., {Kim}, J.-G., \& {Ostriker}, E.~C. 2022, \apjl, 929, L18, \dodoi{10.3847/2041-8213/ac6427}

\bibitem[{{Frerking} {et~al.}(1982){Frerking}, {Langer}, \& {Wilson}}]{Frerking_1982}
{Frerking}, M.~A., {Langer}, W.~D., \& {Wilson}, R.~W. 1982, \apj, 262, 590, \dodoi{10.1086/160451}

\bibitem[{{Gaia Collaboration} {et~al.}(2023){Gaia Collaboration}, {Vallenari}, {Brown}, {Prusti}, {de Bruijne}, {Arenou}, {Babusiaux}, {Biermann}, {Creevey}, {Ducourant}, {Evans}, {Eyer}, {Guerra}, {Hutton}, {Jordi}, {Klioner}, {Lammers}, {Lindegren}, {Luri}, {Mignard}, {Panem}, {Pourbaix}, {Randich}, {Sartoretti}, {Soubiran}, {Tanga}, {Walton}, {Bailer-Jones}, {Bastian}, {Drimmel}, {Jansen}, {Katz}, {Lattanzi}, {van Leeuwen}, {Bakker}, {Cacciari}, {Casta{\~n}eda}, {De Angeli}, {Fabricius}, {Fouesneau}, {Fr{\'e}mat}, {Galluccio}, {Guerrier}, {Heiter}, {Masana}, {Messineo}, {Mowlavi}, {Nicolas}, {Nienartowicz}, {Pailler}, {Panuzzo}, {Riclet}, {Roux}, {Seabroke}, {Sordo}, {Th{\'e}venin}, {Gracia-Abril}, {Portell}, {Teyssier}, {Altmann}, {Andrae}, {Audard}, {Bellas-Velidis}, {Benson}, {Berthier}, {Blomme}, {Burgess}, {Busonero}, {Busso}, {C{\'a}novas}, {Carry}, {Cellino}, {Cheek}, {Clementini}, {Damerdji}, {Davidson}, {de Teodoro}, {Nu{\~n}ez Campos}, {Delchambre}, {Dell'Oro}, {Esquej},
  {Fern{\'a}ndez-Hern{\'a}ndez}, {Fraile}, {Garabato}, {Garc{\'\i}a-Lario}, {Gosset}, {Haigron}, {Halbwachs}, {Hambly}, {Harrison}, {Hern{\'a}ndez}, {Hestroffer}, {Hodgkin}, {Holl}, {Jan{\ss}en}, {Jevardat de Fombelle}, {Jordan}, {Krone-Martins}, {Lanzafame}, {L{\"o}ffler}, {Marchal}, {Marrese}, {Moitinho}, {Muinonen}, {Osborne}, {Pancino}, {Pauwels}, {Recio-Blanco}, {Reyl{\'e}}, {Riello}, {Rimoldini}, {Roegiers}, {Rybizki}, {Sarro}, {Siopis}, {Smith}, {Sozzetti}, {Utrilla}, {van Leeuwen}, {Abbas}, {{\'A}brah{\'a}m}, {Abreu Aramburu}, {Aerts}, {Aguado}, {Ajaj}, {Aldea-Montero}, {Altavilla}, {{\'A}lvarez}, {Alves}, {Anders}, {Anderson}, {Anglada Varela}, {Antoja}, {Baines}, {Baker}, {Balaguer-N{\'u}{\~n}ez}, {Balbinot}, {Balog}, {Barache}, {Barbato}, {Barros}, {Barstow}, {Bartolom{\'e}}, {Bassilana}, {Bauchet}, {Becciani}, {Bellazzini}, {Berihuete}, {Bernet}, {Bertone}, {Bianchi}, {Binnenfeld}, {Blanco-Cuaresma}, {Blazere}, {Boch}, {Bombrun}, {Bossini}, {Bouquillon}, {Bragaglia}, {Bramante}, {Breedt},
  {Bressan}, {Brouillet}, {Brugaletta}, {Bucciarelli}, {Burlacu}, {Butkevich}, {Buzzi}, {Caffau}, {Cancelliere}, {Cantat-Gaudin}, {Carballo}, {Carlucci}, {Carnerero}, {Carrasco}, {Casamiquela}, {Castellani}, {Castro-Ginard}, {Chaoul}, {Charlot}, {Chemin}, {Chiaramida}, {Chiavassa}, {Chornay}, {Comoretto}, {Contursi}, {Cooper}, {Cornez}, {Cowell}, {Crifo}, {Cropper}, {Crosta}, {Crowley}, {Dafonte}, {Dapergolas}, {David}, {David}, {de Laverny}, {De Luise}, {De March}, {De Ridder}, {de Souza}, {de Torres}, {del Peloso}, {del Pozo}, {Delbo}, {Delgado}, {Delisle}, {Demouchy}, {Dharmawardena}, {Di Matteo}, {Diakite}, {Diener}, {Distefano}, {Dolding}, {Edvardsson}, {Enke}, {Fabre}, {Fabrizio}, {Faigler}, {Fedorets}, {Fernique}, {Fienga}, {Figueras}, {Fournier}, {Fouron}, {Fragkoudi}, {Gai}, {Garcia-Gutierrez}, {Garcia-Reinaldos}, {Garc{\'\i}a-Torres}, {Garofalo}, {Gavel}, {Gavras}, {Gerlach}, {Geyer}, {Giacobbe}, {Gilmore}, {Girona}, {Giuffrida}, {Gomel}, {Gomez}, {Gonz{\'a}lez-N{\'u}{\~n}ez},
  {Gonz{\'a}lez-Santamar{\'\i}a}, {Gonz{\'a}lez-Vidal}, {Granvik}, {Guillout}, {Guiraud}, {Guti{\'e}rrez-S{\'a}nchez}, {Guy}, {Hatzidimitriou}, {Hauser}, {Haywood}, {Helmer}, {Helmi}, {Sarmiento}, {Hidalgo}, {Hilger}, {H{\l}adczuk}, {Hobbs}, {Holland}, {Huckle}, {Jardine}, {Jasniewicz}, {Jean-Antoine Piccolo}, {Jim{\'e}nez-Arranz}, {Jorissen}, {Juaristi Campillo}, {Julbe}, {Karbevska}, {Kervella}, {Khanna}, {Kontizas}, {Kordopatis}, {Korn}, {K{\'o}sp{\'a}l}, {Kostrzewa-Rutkowska}, {Kruszy{\'n}ska}, {Kun}, {Laizeau}, {Lambert}, {Lanza}, {Lasne}, {Le Campion}, {Lebreton}, {Lebzelter}, {Leccia}, {Leclerc}, {Lecoeur-Taibi}, {Liao}, {Licata}, {Lindstr{\o}m}, {Lister}, {Livanou}, {Lobel}, {Lorca}, {Loup}, {Madrero Pardo}, {Magdaleno Romeo}, {Managau}, {Mann}, {Manteiga}, {Marchant}, {Marconi}, {Marcos}, {Marcos Santos}, {Mar{\'\i}n Pina}, {Marinoni}, {Marocco}, {Marshall}, {Martin Polo}, {Mart{\'\i}n-Fleitas}, {Marton}, {Mary}, {Masip}, {Massari}, {Mastrobuono-Battisti}, {Mazeh}, {McMillan}, {Messina}, {Michalik},
  {Millar}, {Mints}, {Molina}, {Molinaro}, {Moln{\'a}r}, {Monari}, {Mongui{\'o}}, {Montegriffo}, {Montero}, {Mor}, {Mora}, {Morbidelli}, {Morel}, {Morris}, {Muraveva}, {Murphy}, {Musella}, {Nagy}, {Noval}, {Oca{\~n}a}, {Ogden}, {Ordenovic}, {Osinde}, {Pagani}, {Pagano}, {Palaversa}, {Palicio}, {Pallas-Quintela}, {Panahi}, {Payne-Wardenaar}, {Pe{\~n}alosa Esteller}, {Penttil{\"a}}, {Pichon}, {Piersimoni}, {Pineau}, {Plachy}, {Plum}, {Poggio}, {Pr{\v{s}}a}, {Pulone}, {Racero}, {Ragaini}, {Rainer}, {Raiteri}, {Rambaux}, {Ramos}, {Ramos-Lerate}, {Re Fiorentin}, {Regibo}, {Richards}, {Rios Diaz}, {Ripepi}, {Riva}, {Rix}, {Rixon}, {Robichon}, {Robin}, {Robin}, {Roelens}, {Rogues}, {Rohrbasser}, {Romero-G{\'o}mez}, {Rowell}, {Royer}, {Ruz Mieres}, {Rybicki}, {Sadowski}, {S{\'a}ez N{\'u}{\~n}ez}, {Sagrist{\`a} Sell{\'e}s}, {Sahlmann}, {Salguero}, {Samaras}, {Sanchez Gimenez}, {Sanna}, {Santove{\~n}a}, {Sarasso}, {Schultheis}, {Sciacca}, {Segol}, {Segovia}, {S{\'e}gransan}, {Semeux}, {Shahaf}, {Siddiqui}, {Siebert},
  {Siltala}, {Silvelo}, {Slezak}, {Slezak}, {Smart}, {Snaith}, {Solano}, {Solitro}, {Souami}, {Souchay}, {Spagna}, {Spina}, {Spoto}, {Steele}, {Steidelm{\"u}ller}, {Stephenson}, {S{\"u}veges}, {Surdej}, {Szabados}, {Szegedi-Elek}, {Taris}, {Taylor}, {Teixeira}, {Tolomei}, {Tonello}, {Torra}, {Torra}, {Torralba Elipe}, {Trabucchi}, {Tsounis}, {Turon}, {Ulla}, {Unger}, {Vaillant}, {van Dillen}, {van Reeven}, {Vanel}, {Vecchiato}, {Viala}, {Vicente}, {Voutsinas}, {Weiler}, {Wevers}, {Wyrzykowski}, {Yoldas}, {Yvard}, {Zhao}, {Zorec}, {Zucker}, \& {Zwitter}}]{Gaia_2023}
{Gaia Collaboration}, {Vallenari}, A., {Brown}, A.~G.~A., {et~al.} 2023, \aap, 674, A1, \dodoi{10.1051/0004-6361/202243940}

\bibitem[{{Ginsburg} {et~al.}(2011){Ginsburg}, {Bally}, \& {Williams}}]{Ginsburg_2011}
{Ginsburg}, A., {Bally}, J., \& {Williams}, J.~P. 2011, \mnras, 418, 2121, \dodoi{10.1111/j.1365-2966.2011.19279.x}

\bibitem[{{Hammond} {et~al.}(2023){Hammond}, {Christiaens}, {Price}, {Toci}, {Pinte}, {Juillard}, \& {Garg}}]{Hammond_2023}
{Hammond}, I., {Christiaens}, V., {Price}, D.~J., {et~al.} 2023, \mnras, 522, L51, \dodoi{10.1093/mnrasl/slad027}

\bibitem[{{Hayashi} {et~al.}(1985){Hayashi}, {Nakazawa}, \& {Nakagawa}}]{Hayashi_1985}
{Hayashi}, C., {Nakazawa}, K., \& {Nakagawa}, Y. 1985, in Protostars and Planets II, ed. D.~C. {Black} \& M.~S. {Matthews}, 1100--1153

\bibitem[{{Hirano} {et~al.}(2020){Hirano}, {Tsukamoto}, {Basu}, \& {Machida}}]{Hirano_2020}
{Hirano}, S., {Tsukamoto}, Y., {Basu}, S., \& {Machida}, M.~N. 2020, \apj, 898, 118, \dodoi{10.3847/1538-4357/ab9f9d}

\bibitem[{{Honma} {et~al.}(2014){Honma}, {Akiyama}, {Uemura}, \& {Ikeda}}]{Honma_2014}
{Honma}, M., {Akiyama}, K., {Uemura}, M., \& {Ikeda}, S. 2014, \pasj, 66, 95, \dodoi{10.1093/pasj/psu070}

\bibitem[{Hunter(2007)}]{Hunter_2007}
Hunter, J.~D. 2007, Computing in Science \& Engineering, 9, 90, \dodoi{10.1109/MCSE.2007.55}

\bibitem[{{Jaeger} {et~al.}(1994){Jaeger}, {Mutschke}, {Begemann}, {Dorschner}, \& {Henning}}]{Jaeger_1994}
{Jaeger}, C., {Mutschke}, H., {Begemann}, B., {Dorschner}, J., \& {Henning}, T. 1994, \aap, 292, 641

\bibitem[{{Kepley} {et~al.}(2020){Kepley}, {Tsutsumi}, {Brogan}, {Indebetouw}, {Yoon}, {Mason}, \& {Donovan Meyer}}]{Kepley_2020}
{Kepley}, A.~A., {Tsutsumi}, T., {Brogan}, C.~L., {et~al.} 2020, \pasp, 132, 024505, \dodoi{10.1088/1538-3873/ab5e14}

\bibitem[{{Keppler} {et~al.}(2019){Keppler}, {Teague}, {Bae}, {Benisty}, {Henning}, {van Boekel}, {Chapillon}, {Pinilla}, {Williams}, {Bertrang}, {Facchini}, {Flock}, {Ginski}, {Juhasz}, {Klahr}, {Liu}, {M{\"u}ller}, {P{\'e}rez}, {Pohl}, {Rosotti}, {Samland}, \& {Semenov}}]{Keppler_2019}
{Keppler}, M., {Teague}, R., {Bae}, J., {et~al.} 2019, \aap, 625, A118, \dodoi{10.1051/0004-6361/201935034}

\bibitem[{{Kuramochi} {et~al.}(2018){Kuramochi}, {Akiyama}, {Ikeda}, {Tazaki}, {Fish}, {Pu}, {Asada}, \& {Honma}}]{Kuramochi_2018}
{Kuramochi}, K., {Akiyama}, K., {Ikeda}, S., {et~al.} 2018, \apj, 858, 56, \dodoi{10.3847/1538-4357/aab6b5}

\bibitem[{{Lacy} {et~al.}(2020){Lacy}, {Baum}, {Chandler}, {Chatterjee}, {Clarke}, {Deustua}, {English}, {Farnes}, {Gaensler}, {Gugliucci}, {Hallinan}, {Kent}, {Kimball}, {Law}, {Lazio}, {Marvil}, {Mao}, {Medlin}, {Mooley}, {Murphy}, {Myers}, {Osten}, {Richards}, {Rosolowsky}, {Rudnick}, {Schinzel}, {Sivakoff}, {Sjouwerman}, {Taylor}, {White}, {Wrobel}, {Andernach}, {Beasley}, {Berger}, {Bhatnager}, {Birkinshaw}, {Bower}, {Brandt}, {Brown}, {Burke-Spolaor}, {Butler}, {Comerford}, {Demorest}, {Fu}, {Giacintucci}, {Golap}, {G{\"u}th}, {Hales}, {Hiriart}, {Hodge}, {Horesh}, {Ivezi{\'c}}, {Jarvis}, {Kamble}, {Kassim}, {Liu}, {Loinard}, {Lyons}, {Masters}, {Mezcua}, {Moellenbrock}, {Mroczkowski}, {Nyland}, {O'Dea}, {O'Sullivan}, {Peters}, {Radford}, {Rao}, {Robnett}, {Salcido}, {Shen}, {Sobotka}, {Witz}, {Vaccari}, {van Weeren}, {Vargas}, {Williams}, \& {Yoon}}]{Lacy_2020}
{Lacy}, M., {Baum}, S.~A., {Chandler}, C.~J., {et~al.} 2020, \pasp, 132, 035001, \dodoi{10.1088/1538-3873/ab63eb}

\bibitem[{{Luhman}(2025)}]{Luhman_2025}
{Luhman}, K.~L. 2025, \aj, 169, 179, \dodoi{10.3847/1538-3881/adb0cd}

\bibitem[{{Machida} {et~al.}(2020){Machida}, {Hirano}, \& {Kitta}}]{Machida_2020}
{Machida}, M.~N., {Hirano}, S., \& {Kitta}, H. 2020, \mnras, 491, 2180, \dodoi{10.1093/mnras/stz3159}

\bibitem[{{Marino} {et~al.}(2015){Marino}, {Perez}, \& {Casassus}}]{Marino_2015}
{Marino}, S., {Perez}, S., \& {Casassus}, S. 2015, \apjl, 798, L44, \dodoi{10.1088/2041-8205/798/2/L44}

\bibitem[{{Nagasawa} {et~al.}(2008){Nagasawa}, {Ida}, \& {Bessho}}]{Nagasawa_2008}
{Nagasawa}, M., {Ida}, S., \& {Bessho}, T. 2008, \apj, 678, 498, \dodoi{10.1086/529369}

\bibitem[{{Nakazato} \& {Ikeda}(2020)}]{Nakazato_2020}
{Nakazato}, T., \& {Ikeda}, S. 2020, {PRIISM: Python module for Radio Interferometry Imaging with Sparse Modeling}, Astrophysics Source Code Library, record ascl:2006.002

\bibitem[{Nakazato {et~al.}(2020)Nakazato, Ikeda, Kosugi, \& Honma}]{Nakazato_2020b}
Nakazato, T., Ikeda, S., Kosugi, G., \& Honma, M. 2020, in Millimeter, Submillimeter, and Far-Infrared Detectors and Instrumentation for Astronomy X, ed. J.~Zmuidzinas \& J.-R. Gao, Vol. 11453, International Society for Optics and Photonics (SPIE), 114532V, \dodoi{10.1117/12.2560904}

\bibitem[{{Nealon} {et~al.}(2025){Nealon}, {Smallwood}, {Aly}, {Winter}, {Longarini}, {Cuello}, {Veras}, \& {Alexander}}]{Nealon_2025}
{Nealon}, R., {Smallwood}, J.~L., {Aly}, H., {et~al.} 2025, \mnras, 540, L84, \dodoi{10.1093/mnrasl/slaf032}

\bibitem[{{Orihara} {et~al.}(2023){Orihara}, {Momose}, {Muto}, {Hashimoto}, {Liu}, {Tsukagoshi}, {Kudo}, {Takahashi}, {Yang}, {Hasegawa}, {Dong}, {Konishi}, \& {Akiyama}}]{Orihara_2023}
{Orihara}, R., {Momose}, M., {Muto}, T., {et~al.} 2023, \pasj, 75, 424, \dodoi{10.1093/pasj/psad009}

\bibitem[{{Pascucci} {et~al.}(2016){Pascucci}, {Testi}, {Herczeg}, {Long}, {Manara}, {Hendler}, {Mulders}, {Krijt}, {Ciesla}, {Henning}, {Mohanty}, {Drabek-Maunder}, {Apai}, {Sz{\H{u}}cs}, {Sacco}, \& {Olofsson}}]{Pascucci_2016}
{Pascucci}, I., {Testi}, L., {Herczeg}, G.~J., {et~al.} 2016, \apj, 831, 125, \dodoi{10.3847/0004-637X/831/2/125}

\bibitem[{{Perez} {et~al.}(2015){Perez}, {Casassus}, {M{\'e}nard}, {Roman}, {van der Plas}, {Cieza}, {Pinte}, {Christiaens}, \& {Hales}}]{Peres_2015}
{Perez}, S., {Casassus}, S., {M{\'e}nard}, F., {et~al.} 2015, \apj, 798, 85, \dodoi{10.1088/0004-637X/798/2/85}

\bibitem[{{Pinte} {et~al.}(2018){Pinte}, {Price}, {M{\'e}nard}, {Duch{\^e}ne}, {Dent}, {Hill}, {de Gregorio-Monsalvo}, {Hales}, \& {Mentiplay}}]{Pinte_2018}
{Pinte}, C., {Price}, D.~J., {M{\'e}nard}, F., {et~al.} 2018, \apjl, 860, L13, \dodoi{10.3847/2041-8213/aac6dc}

\bibitem[{{Rau} \& {Cornwell}(2011)}]{Rau_2011}
{Rau}, U., \& {Cornwell}, T.~J. 2011, \aap, 532, A71, \dodoi{10.1051/0004-6361/201117104}

\bibitem[{{Rota} {et~al.}(2024){Rota}, {Meijerhof}, {van der Marel}, {Francis}, {van der Tak}, \& {Sellek}}]{Rota_2024}
{Rota}, A.~A., {Meijerhof}, J.~D., {van der Marel}, N., {et~al.} 2024, \aap, 684, A134, \dodoi{10.1051/0004-6361/202348387}

\bibitem[{{Schwab}(1984)}]{Schwab_1984}
{Schwab}, F.~R. 1984, \aj, 89, 1076, \dodoi{10.1086/113605}

\bibitem[{{Shoshi} {et~al.}(2025){Shoshi}, {Yamaguchi}, {Muto}, {Hirano}, {Kawabe}, {Tsukagoshi}, \& {Machida}}]{Shoshi_2025}
{Shoshi}, A., {Yamaguchi}, M., {Muto}, T., {et~al.} 2025, \pasj, \dodoi{10.1093/pasj/psaf026}

\bibitem[{{Shoshi} {et~al.}(2024){Shoshi}, {Harada}, {Tokuda}, {Kawasaki}, {Yamasaki}, {Sato}, {Omura}, {Yamaguchi}, {Tachihara}, \& {Machida}}]{Shoshi_2024}
{Shoshi}, A., {Harada}, N., {Tokuda}, K., {et~al.} 2024, \apj, 961, 228, \dodoi{10.3847/1538-4357/ad12b5}

\bibitem[{{Shu} {et~al.}(1987){Shu}, {Adams}, \& {Lizano}}]{Shu_1987}
{Shu}, F.~H., {Adams}, F.~C., \& {Lizano}, S. 1987, \araa, 25, 23, \dodoi{10.1146/annurev.aa.25.090187.000323}

\bibitem[{{Testi} {et~al.}(2022){Testi}, {Natta}, {Manara}, {de Gregorio Monsalvo}, {Lodato}, {Lopez}, {Muzic}, {Pascucci}, {Sanchis}, {Miranda}, {Scholz}, {De Simone}, \& {Williams}}]{Testi_2022}
{Testi}, L., {Natta}, A., {Manara}, C.~F., {et~al.} 2022, \aap, 663, A98, \dodoi{10.1051/0004-6361/202141380}

\bibitem[{{van der Marel} {et~al.}(2016){van der Marel}, {van Dishoeck}, {Bruderer}, {Andrews}, {Pontoppidan}, {Herczeg}, {van Kempen}, \& {Miotello}}]{vanderMarel_2016}
{van der Marel}, N., {van Dishoeck}, E.~F., {Bruderer}, S., {et~al.} 2016, \aap, 585, A58, \dodoi{10.1051/0004-6361/201526988}

\bibitem[{{Virtanen} {et~al.}(2020){Virtanen}, {Gommers}, {Oliphant}, {Haberland}, {Reddy}, {Cournapeau}, {Burovski}, {Peterson}, {Weckesser}, {Bright}, {van der Walt}, {Brett}, {Wilson}, {Millman}, {Mayorov}, {Nelson}, {Jones}, {Kern}, {Larson}, {Carey}, {Polat}, {Feng}, {Moore}, {VanderPlas}, {Laxalde}, {Perktold}, {Cimrman}, {Henriksen}, {Quintero}, {Harris}, {Archibald}, {Ribeiro}, {Pedregosa}, {van Mulbregt}, \& {SciPy 1. 0 Contributors}}]{Virtanen_2020}
{Virtanen}, P., {Gommers}, R., {Oliphant}, T.~E., {et~al.} 2020, Nature Methods, 17, 261, \dodoi{10.1038/s41592-019-0686-2}

\bibitem[{{Yamaguchi} {et~al.}(2021){Yamaguchi}, {Tsukagoshi}, {Muto}, {Nomura}, {Nakazato}, {Ikeda}, {Tamura}, \& {Kawabe}}]{Yamaguchi_2021}
{Yamaguchi}, M., {Tsukagoshi}, T., {Muto}, T., {et~al.} 2021, \apj, 923, 121, \dodoi{10.3847/1538-4357/ac2bfd}

\bibitem[{{Yamaguchi} {et~al.}(2024){Yamaguchi}, {Muto}, {Tsukagoshi}, {Nomura}, {Hirano}, {Nakazato}, {Ikeda}, {Tamura}, \& {Kawabe}}]{Yamaguchi_2024}
{Yamaguchi}, M., {Muto}, T., {Tsukagoshi}, T., {et~al.} 2024, \pasj, 76, 437, \dodoi{10.1093/pasj/psae022}

\end{thebibliography}
\bibliographystyle{aasjournal}
\end{document}